\documentclass[12pt]{article}
\usepackage{amsmath}
\usepackage{graphicx}
\usepackage{enumerate}
\usepackage{tikz}
\usepackage[numbers, sort, comma, square]{natbib}
\usepackage{url} 
\usepackage{moreverb,mathtools,amsfonts,amssymb,bbold,float}
\usepackage{caption}
\usepackage{amsmath}
\usepackage{graphicx,psfrag,epsf}
\usepackage{subcaption}
\usepackage{tikz}
\usetikzlibrary{positioning}
\usepackage{bbm}
\usepackage{soul}
\usepackage{tablefootnote}
\usepackage{cancel}
\usepackage{ulem}
\usepackage{longtable}
\usepackage{lipsum}
\usepackage{cite}
\usepackage{listings}
\usepackage{verbatim}
\usepackage{color} 
\usepackage{enumitem} 
\usepackage{makecell}
\usepackage{multicol}
\usepackage{verbatim}
\usepackage{amsthm}
\usepackage{amstext} 
\usepackage{array} 
\usepackage{algorithm}
\usepackage{verbatim}
\usepackage{chngcntr}
\usepackage{apptools}
\AtAppendix{\counterwithin{definition}{section}}
\AtAppendix{\counterwithin{theorem}{section}}
\newcolumntype{L}{>{$}l<{$}} 
\usetikzlibrary{calc,trees,positioning,arrows,chains,shapes.geometric,%
    decorations.pathreplacing,decorations.pathmorphing,shapes,%
    matrix,shapes.symbols}
    \tikzset{
>=stealth',
  punktchain/.style={
    rectangle, 
    rounded corners, 
    draw=black, very thick,
    text width=10em, 
    minimum height=3em, 
    text centered, 
    on chain},
  line/.style={draw, thick, <-},
  element/.style={
    tape,
    top color=white,
    bottom color=blue!50!black!60!,
    minimum width=8em,
    draw=blue!40!black!90, very thick,
    text width=10em, 
    minimum height=3.5em, 
    text centered, 
    on chain},
  every join/.style={->, thick,shorten >=1pt},
  decoration={brace},
  tuborg/.style={decorate},
  tubnode/.style={midway, right=2pt},
}
\newtheorem{theorem}{Theorem}

\newtheorem{definition}{Definition}
\usepackage{authblk}
\usepackage{fancybox}
\usepackage{tikz}
\usepackage{graphicx,psfrag,epsf}
\usepackage{pgf,tikz}
\usetikzlibrary{decorations.pathreplacing,angles,quotes}
\usetikzlibrary{shapes.geometric, arrows,positioning}
\tikzstyle{line}=[draw]
\usepackage{booktabs}
\usepackage{graphicx}
\usepackage{tabularx}
\usepackage{arydshln}
\usepackage{hyperref}
\renewcommand{\baselinestretch}{1.5}
\usepackage{extarrows}%
\usepackage{multirow}
\newcommand\norm[1]{\left\lVert#1\right\rVert}
\newcommand{\blind}{1}

\begin{document}

\def\spacingset#1{\renewcommand{\baselinestretch}%
{#1}\small\normalsize} \spacingset{1}


\if1\blind
{
  \title{\bf Integrating complex selection rules into the latent overlapping group Lasso for the construction of coherent prediction models
  }
  \author[1]{{\small Guanbo Wang*}}
  \author[2]{{\small Sylvie Perreault}}
  \author[3,4]{{\small Robert W. Platt}}
  \author[5,6]{{\small Rui Wang}}
  \author[7]{{\small Marc Dorais}}
  \author[2,8,3]{{\small Mireille E. Schnitzer}}
    
  \affil[1]{\small CAUSALab, Department of Epidemiology, Harvard T. H. Chan School of Public Health, Harvard University, Boston, MA, USA}
  \affil[2]{\small Facult\'e de pharmacie, Universit\'e de Montr\'eal, Montr\'eal, QC, Canada}
  \affil[3]{\small Department of Epidemiology, Biostatistics and Occupational Health, McGill University, Montr\'eal, QC, Canada}
  \affil[4]{\small Department of Pediatrics, McGill University, Montr\'eal, QC, Canada}
  \affil[5]{\small Harvard Pilgrim Health Care Institute and Harvard Medical School, Boston, MA, USA}
  \affil[6]{\small Department of Biostatistics, Harvard T. H. Chan School of Public Health, Harvard University, Boston, MA, USA}
  \affil[7]{\small StatSciences Inc., Notre-Dame-de-l'Île-Perrot, Quebec, Canada}
  \affil[8]{\small D\'epartement de m\'edecine sociale et pr\'eventive,  Universit\'e de Montr\'eal, Montr\'eal, QC, Canada}
  \renewcommand\Authands{, }
  \date{}
  \maketitle
} \fi

\if0\blind
{
  \bigskip
  \bigskip
  \bigskip
  \begin{center}
    {\LARGE\bf Integrating complex selection rules into the latent overlapping group Lasso for constructing coherent prediction models}
\end{center}
  \medskip
} \fi

\bigskip
\begin{abstract}
\noindent
The construction of coherent prediction models holds great importance in medical research as such models enable health researchers to gain deeper insights into disease epidemiology and clinicians to identify patients at higher risk of adverse outcomes. One commonly employed approach to developing prediction models is variable selection through penalized regression techniques. Integrating natural variable structures into this process not only enhances model interpretability but can also 
boost prediction accuracy. However, a challenge lies in determining how to effectively integrate potentially complex selection dependencies into the penalized regression. In this work, we demonstrate how to represent selection dependencies mathematically, provide algorithms for deriving the complete set of potential models, and offer a structured approach for integrating complex rules into variable selection through the latent overlapping group Lasso. To illustrate our methodology, we applied these techniques to construct a coherent prediction model for major bleeding in hypertensive patients recently hospitalized for atrial fibrillation and subsequently prescribed oral anticoagulants. In this application, we account for a proxy of anticoagulant adherence and its interaction with dosage and the type of oral anticoagulants in addition to drug-drug interactions.
\end{abstract}
\noindent%
{\it Keywords:} coherent prediction models, grouping structure, interpretability, selection rules, structured sparsity pattern, the latent overlapping group lasso,
\vfill

\newpage
\spacingset{1.5} 
\section{Introduction}
In medical research, constructing coherent prediction models for an outcome can help health researchers develop a better understanding of disease epidemiology, i.e., which factors are associated with a higher risk of the disease of interest and the potential contribution of each one to the overall prediction  \citep{schooling2018clarifying}. The ensuing prediction models can then assist clinicians in identifying patients who are at higher risk of the outcome, in order to potentially provide different care or more intensive follow-up. Constructing coherent prediction models can also be regarded as model exploration for causal hypotheses: the predictors being identified can be potentially investigated in future causal analyses and randomized controlled trials \citep{shmueli2010explain,kalisch2007estimating,shortreed2017outcome,siddique2019causal,Wang2020Estimating,liu2022modeling,bouchard2022predictive,Wang2023evaluating,Wang2023Review}. 

With the increasing use of administrative claims and electronic health records (EHR), higher-dimensional patient information is accessible, which may allow for the development of more accurate prediction models. When one wants a prediction model that can be understood by users for both credibility and usefulness in clinical decision-making, constructing a non-parametric prediction model may be less preferable because black box models are difficult to interpret, and prediction accuracy is not the only goal that we pursue. To strike a balance between prediction accuracy and interpretability, a vast literature of variable selection techniques under generalized linear models has emerged \citep{tibshirani1996regression,zou2006adaptive,yuan2006model,bhatnagar2020sparse,breheny2009penalized,mairal2010network,jacob2009group}.

Selection dependencies are often inherent or desirable in variable selection. For example, binary indicators representing non-reference categories of a categorical variable should be selected collectively. As a second example, the inclusion of a variable (for instance, an interaction term) in a model can depend on whether other variables (for instance, main terms) are also included. Such selection dependencies place a constraint on the allowable candidate models; that is, they limit which combinations of covariables can be present in the prediction model. Models that satisfy these selection dependencies are coherent by construction and thus ensure the interpretability of the results.

Recently, \citet{wang2023general} developed a framework for structured variable selection that allows analysts to represent any combination of selection dependencies, called ``selection rules'', using mathematical language, and then construct the corresponding ``selection dictionary'', which is a set that contains all subsets of candidate variables that respect the selection rule. Models with covariate sets belonging to this selection dictionary are defined as coherent with the selection rule. We can then perform model selection by fitting models with covariates corresponding to each set of candidate variables in the selection dictionary, and select the model that has the best performance in some sense (for example, the lowest cross-validated risk under some loss function).

When the size of the selection dictionary is large, applying this exhaustive method is computationally inefficient. Another solution to accommodate complex selection rules to arrive at a coherent prediction model via variable selection is penalized regression, where the selection dependency is achieved by specifying a ``grouping structure'', that is, assigning (possibly overlapping) groups of coefficients of covariates in the penalty term. However, when dealing with complex variable dependencies, the relationship between grouping structure and selection rules becomes less clear. In other words, determining how to appropriately group variables together in order to adhere to the selection rules remains a challenging task.

In this study, we develop practical guidance for structured variable selection, broadly applicable across various domains. Based on the theoretical framework developed by \citet{wang2023general}, we develop and compare two distinct algorithms to compute the selection dictionary. The latent overlapping group Lasso (LOGL) \citep{obozinski2011variable} is the most versatile penalized regression for selection rules, and we take it as an example to show how to incorporate selection rules into variable selection using a penalized regression. Specifically, under the LOGL, we derive the sufficient and necessary conditions for a grouping structure to respect a given selection rule and offer practical roadmaps for identifying grouping structures for commonly used selection rules. We show in the web materials that such conditions can be generalized to the overlapping group Lasso and general penalized regressions. 

Throughout the paper, we demonstrate the developed techniques through a real-world example. Patients with atrial fibrillation who initiated oral anticoagulants (OACs) have risks of major bleeding. Constructing a coherent prediction model for major bleeding would help health researchers identify important predictors for major bleeding. We applied our approach to develop a prediction model for major bleeding incorporating 10 selection rules relating to the nested structure of the categorical variables and background knowledge, with the goal of producing a model with face validity in the field. In this application, we consider different OAC drugs, adherence to drug prescriptions, and drug-drug interactions as candidate predictors, showcasing the applicability of our approach in a complex setting.

The remainder of the paper is organized as follows. Section \ref{data_example} introduces the motivating data example. Section \ref{sec:S-S} starts with a brief introduction to the structured variable selection framework, focusing on formulating the selection dependencies, followed by a demonstration of the application of the framework in developing the selection rules and related dictionaries for the motivating example. In Section \ref{sec:SD-GS}, we first review the LOGL, show how to group variables in the penalty term to respect selection rules, and then implement the methods for the motivating example. The results of the data analyses for the motivating example are given in section \ref{results}, followed by a discussion in section \ref{discussion}. 

\section{Predicting major bleeding in patients with atrial fibrillation prescribed anticoagulants}\label{sec:MotivatingData}
\label{data_example}
Atrial fibrillation is a disease characterized by an irregular heartbeat, due to electrical signal disturbances of the heart. Patients with this condition have a higher risk of stroke, heart failure, and other cardiovascular complications \citep{lip2007management}. To decrease the occurrence of stroke, most patients have to take anticoagulants long-term \citep{yamashiro2019adequate}. Historically, warfarin (vitamin K antagonist) was the mainstay anticoagulant for non-valvular atrial fibrillation \citep{friberg2012net}. However, close monitoring of the International Normalized Ratio (INR) and frequent dose adjustments due to numerous drug interactions are required when taking warfarin \citep{connolly2008benefit}. In recent decades, direct oral anticoagulants (DOACs), including rivaroxaban, dabigatran, and apixaban, became available to patients with non-valvular atrial fibrillation. As alternatives to warfarin, they do not require such routine monitoring and are given as fixed doses based on patient characteristics. Nevertheless, non-compliance \citep{obamiro2016summary, garkina2016compliance}, different dose levels, drug interactions \citep{raccah2018drug, burn2018direct} and contraindications \citep{schnitzer2020tutorial} complicate the usage of anticoagulants. Patients taking different anticoagulants with various doses and adherence patterns may have different risks of major bleeding. Beyond that, many known or unknown factors may also contribute to variability in outcomes. Therefore, before measuring the effects of anticoagulants or predicting the risk of major bleeding, we are interested in investigating the predictors that are associated with or predictive of major bleeding through modeling \citep{tripodi2018and, claxton2018new}.

We use the dataset from \citet{qazi2021predicting}, compiled from a subset of the R\'egie de l'Assurance Maladie du Qu\'ebec (RAMQ) drug and medical services database linked with the Med-Echo hospitalization database using encrypted patient healthcare insurance numbers \citep{tamblyn1995use,perreault2020oral,eguale2010enhancing,wilchesky2004validation}. They identified patients hospitalized for any cause and discharged alive in the community from 2011 to 2017 with a primary or secondary diagnosis of atrial fibrillation. Cohort entry (index date) was defined as the time of the first OAC claim. 

We are particularly interested in assessing if adherence (high/low) to the prescribed OAC is useful in predicting the risk of major bleeding \citep{perreault2019anticoagulants}. However, the history of OAC usage is not available because the cohort consists of patients who used OACs for the first time. We thus used the history of hypertension drug usage, which is also taken chronically, as a proxy for adherence to OACs \citep{sabate2003adherence}. We thus limited the cohort to patients with a previous diagnosis of hypertension who were prescribed at least one hypertension drug in the three months prior to the index date. The complete inclusion and exclusion criteria with patient totals are shown in web material 1 Section \ref{exclusion}.

The outcome is defined as incident major bleeding within 1 year of follow-up. Validated ICD-9 and ICD-10 codes for the outcome are given in web material 2
\citep{villines2015comparison,yao2016effectiveness,lauffenburger2015effectiveness,maura2015comparison,graham2015cardiovascular,go2017outcomes}. We screened the dataset and selected variables that were either known to be predictive of major bleeding or of particular clinical interest \citep{qazi2021predicting,landefeld1989major,roldan2013has,chao2018major}, and the availability of data.
When available, we included the variables in the HAS-BLED (Hypertension, Abnormal Renal/Liver Function, Stroke, Bleeding History or Predisposition, Labile INR, Elderly, Drugs/Alcohol Concomitantly) chart \citep{pisters2010novel} and known risk factors of bleeding when taking OACs \citep{lane2012use}; the definitions of these variables are given in web material 2. The potential predictors in our analysis included age, sex, CHA$_{2}$DS$_{2}$-VASc score, comorbidities within 3 years before cohort entry, OAC type used at cohort entry, concomitant medication usage within 2 weeks before cohort entry, and the drug-drug interactions between OAC type and concomitant medications. Definitions of the main terms (binary) variables related to OAC usage are given in Table \ref{tab:definitions_OAC}. \texttt{DOAC} indicates that a patient was prescribed a DOAC, with warfarin as the alternative. We also included indicators of Apixaban and Dabigatran prescriptions, with Rivaroxaban being the reference DOAC. DOACs can also be prescribed at high or low-doses. Our variable \texttt{High-dose-DOAC} indicates that a patient was receiving a high-dose of a DOAC, where \texttt{High-dose-DOAC}$=0$ means that the patient was either receiving low-dose-DOAC or warfarin. Note that the dosing of warfarin is individualized and not considered in this analysis. See web material 1 Section \ref{app:corr} for a full correspondence between all possible drugs taken by a patient and the variable coding. Adherence was calculated as the number of days of dispensed hypertension drugs divided by the duration of the prescription period prior to the index date \citep{tajeu2019antihypertensive}. The variable \texttt{High-adherence} was defined as $\geqslant80\%$ adherence to hypertension drugs. 
\begin{table}[]
    \centering
    \begin{tabular}{p{3.25cm}p{12.5cm}}
        \hline
    Name & Definition \\
    \hline\hline
    \texttt{DOAC}     &  1 if the patient was prescribed DOAC, 0 if the patient was prescribed warfarin at cohort entry\\
    \texttt{Apixaban}     & 1 if the patient was prescribed Apixaban at cohort entry, 0 otherwise\\
    \texttt{Dabigatran} & 1 if the patient was prescribed Dabigatran at cohort entry, 0 otherwise\\
    \texttt{High-dose-DOAC}  & 1 if the patient was prescribed high-dose-DOACs (Apixaban, Dabigatran, Rivaroxaban) at cohort entry, 0 otherwise\\
    \texttt{High-adherence}* & 1 if the patient’s adherence level was greater than or equal to 0.8 \\
    \hline
    \end{tabular}
    {\footnotesize *For the complete definition of adherence, see web material 2.}\\
    \caption{Definitions of variables related to OAC usage}
    \label{tab:definitions_OAC}
\end{table}

\section{From selection rules to selection dictionaries}\label{sec:S-S}
The covariates described earlier demonstrate complex patterns and relationships. For example, it is difficult to interpret models with interactions but not their main terms. In addition, some covariates are ``nested'' in others; the interpretations of the coefficients of those covariates are only of interest if their ``parent'' covariates are also in the model. For instance, \texttt{High-dose-DOAC} is nested in \texttt{DOAC}; it can take a value of one only when \texttt{DOAC} takes a value of one. Therefore, the coefficient of \texttt{High-dose-DOAC} is not easily interpretable or is not of interest without including \texttt{DOAC} in the model. 
Such covariate structures can be translated into selection rules with the goal of integrating them into the variable selection. However, the selection rules are often presented in a verbal format (e.g., if \texttt{High-dose-DOAC} is in the model, then  \texttt{DOAC} must also be in the model) without precise mathematical definitions, which can impede formal integration into algorithmic variable selection. 

In Section \ref{sec:GeneralSelectionRule}, we first review some concepts of selection rules and dictionaries developed by \citet{wang2023general} that are relevant to this application. Then, moving on to Section \ref{sec:CompoundSelectionRule}, we shift our attention back to the running example and illustrate in detail how to transform the aforementioned covariate structure into selection rules. Following this, we provide algorithms to derive the selection dictionary based on the selection rules. The dictionary serves as a tool to guide the grouping of variables within the context of a penalized regression approach.
\subsection{General selection rules and dictionaries}\label{sec:GeneralSelectionRule}
Denote the set of candidate covariates by $\mathbb{V}$. A \textit{selection rule} $\mathfrak{r}$ of $\mathbb{V}$ is defined as selection dependencies among the variables in $\mathbb{V}$, and the \textit{selection dictionary} $\mathbb{D}_{\mathfrak{r}}$ is a unique set that contains all subsets of $\mathbb{V}$ that respect the selection rule $\mathfrak{r}$. We say that $\mathbb{D}_{\mathfrak{r}}$ is congruent to $\mathfrak{r}$, denoted $\mathbb{D}_{\mathfrak{r}}\cong\mathfrak{r}$. For example, the power set $\mathcal{P}(\mathbb{V})$, which is the set of all subsets of $\mathbb{V}$, is the dictionary in the absence of selection rules. 

Denote a subset of $\mathbb{V}$ by $\mathbb{F}$, and a set of numbers $\mathbb{C}$ where $|\mathbb{F}|\geqslant \max(\mathbb{C})$.
To mathematically express selection rules, we first define a \textit{unit rule} $\mathfrak{u}_{\mathbb{C}}(\mathbb{F})$ as a special selection rule, which imposes ``select a number of variables from  $\mathbb{F}$, where the number can correspond to any value in $\mathbb{C}$''. For example, $\mathfrak{u}_{\{1, 3\}}(\mathbb{F})$ translates to ``select one or three variables in $\mathbb{F}$''. Its corresponding \textit{unit dictionary}, i.e. the selection dictionary corresponding to the unit rule, can be shown to be $\{\mathbb{a}\cup\mathbb{b}: \forall\mathbb{a}\subseteq \mathbb{F} \text{ }s.t.\text{ } |\mathbb{a}|\in \mathbb{C}, \forall \mathbb{b}\subseteq\mathbb{V}\setminus\mathbb{F}\}$. We previously showed that all possible selection rules can be expressed by either unit rules or operations on unit rules using the operators $\land$ and $\lor$. The definitions of some operations and the corresponding selection dictionaries are given in Table \ref{tab:Loperations}.
\begin{table}[htbp]
    \begin{center}
    \begin{tabular}{lp{6cm}p{5.5cm}}\hline
     \textbf{Operation} & \textbf{Interpretation}  & \textbf{Selection dictionary} 
     \\\hline\hline
     $\neg \mathfrak{r}_1$ & do not respect $\mathfrak{r}_1$ & $\mathcal{P}(\mathbb{V})\setminus\mathbb{D}_{\mathfrak{r}_{1}}$\\
        $\mathfrak{r}_{1}\land \mathfrak{r}_{2}$     &
        respect both $\{\mathfrak{r}_{1}$, $\mathfrak{r}_{2}\}$  &$\mathbb{D}_{\mathfrak{r}_{1}}\cap\mathbb{D}_{\mathfrak{r}_{2}}$\\
        $\mathfrak{r}_{1}\lor \mathfrak{r}_{2}$     &  respect at least one of  $\{\mathfrak{r}_{1}$, $\mathfrak{r}_{2}\}$ &
        $\mathbb{D}_{\mathfrak{r}_{1}}\cup\mathbb{D}_{\mathfrak{r}_{2}}$\\
  $\mathfrak{r}_{1}\rightarrow \mathfrak{r}_{2}$ & respect $\mathfrak{r}_{2}$ if $\mathfrak{r}_{1}$ is respected &
  $(\mathcal{P}(\mathbb{V})\setminus\mathbb{D}_{\mathfrak{r}_{1}})\cup(\mathbb{D}_{\mathfrak{r}_{1}}\cap\mathbb{D}_{\mathfrak{r}_{2}})$
  \\\hline
    \end{tabular}\\
    \end{center}
    \caption{Results from \citep{wang2023general}: Operations for selection rules and the resulting selection dictionaries. $\mathbb{D}_{\mathfrak{r}_{1}}$ and $\mathbb{D}_{\mathfrak{r}_{2}}$ are the selection dictionaries of $\mathfrak{r}_{1}$ and $\mathfrak{r}_{2}$, respectively.}
    \label{tab:Loperations}
\end{table}
Note that the if-then rule $\mathfrak{r}_{1}\rightarrow \mathfrak{r}_{2}$ can be re-expressed by $(\mathfrak{r}_1\land\mathfrak{r}_2)\lor\mathfrak{r}_2\lor(\neg \mathfrak{r}_1 \land \neg \mathfrak{r}_2)$. 

Theoretically, with the above-defined tools, we can proceed to express any selection rule as a mathematical object and derive its selection dictionary. 
\subsection{Selection rules for the application}\label{sec:CompoundSelectionRule}
Now, we apply these concepts in the motivating example.
Given covariates and the interactions of interest described in Section \ref{sec:MotivatingData}, we consider three types of selection rules, 1) selection rules for OAC usage, which are set mainly to maintain the interpretability of the coefficients of interest, 2) selection rules for drug-drug interactions, where we adopt the principle of strong heredity \citep{haris2016convex} for all interactions: if the interaction is selected, its main terms must be selected, and 3) selection rules for pre-selected variables, based on prior knowledge. The 10 individual selection rules are presented in Table \ref{SRs}. The selection rule for the analysis is the combination of these 10 individual selection rules through the $\land$ operation. The rationale for each rule follows.

\begin{table}[htbp]
    \begin{center}
    \begin{tabular}{ll}
    \hline
     \textbf{\#} & \textbf{Selection Rule} 
     \\\hline\hline
     & \underline{\textit{Selection rules for OAC usage}}\\
 1.1 &  $\mathfrak{u}_{\{1\}}(\{\texttt{High-dose-DOAC}\})\rightarrow\mathfrak{u}_{\{1\}}(\{\texttt{DOAC}\})$ \\  
  1.2 &    $\mathfrak{u}_{\{1\}}(\{\texttt{Apixaban}\})\rightarrow\mathfrak{u}_{\{1\}}(\{\texttt{DOAC}\})$   \\ 
  1.3 &    $\mathfrak{u}_{\{1\}}(\{\texttt{Dabigatran}\})\rightarrow\mathfrak{u}_{\{1\}}(\{\texttt{DOAC}\})$     \\
  1.4 &    $\mathfrak{u}_{\{1\}}(\{\texttt{DOAC}\times\texttt{High-adherence}\})\rightarrow\mathfrak{u}_{\{2\}}(\{\texttt{DOAC, High-adherence}\})$  \\
  1.5 &    $\mathfrak{u}_{\{1\}}(\{\texttt{High-dose-DOAC}\times\texttt{High-adherence}\})\rightarrow\mathfrak{u}_{\{4\}}(\{\texttt{DOAC}, \texttt{High-adherence}$,\\
  & \hspace{0.2cm} \texttt{High-dose-DOAC, DOAC} $\times$ \texttt{High-adherence}\}) \\
  &\underline{\textit{Selection rules for drug-drug interaction}}\\
2.1 &    $\mathfrak{u}_{\{1\}}(\{\texttt{DOAC}\times\texttt{Antiplatelets}\})\rightarrow\mathfrak{u}_{\{2\}}(\{\texttt{DOAC, Antiplatelets}\})$  \\
2.2 &    $\mathfrak{u}_{\{1\}}(\{\texttt{DOAC}\times\texttt{NSAIDs}\})\rightarrow\mathfrak{u}_{\{2\}}(\{\texttt{DOAC, NSAIDs}\})$  \\
2.3 &    $\mathfrak{u}_{\{1\}}(\{\texttt{DOAC}\times\texttt{Antidepressants}\})\rightarrow\mathfrak{u}_{\{2\}}(\{\texttt{DOAC, Antidepressants}\})$  \\
2.4 &    $\mathfrak{u}_{\{1\}}(\{\texttt{DOAC}\times\texttt{PPIs}\})\rightarrow\mathfrak{u}_{\{2\}}(\{\texttt{DOAC, PPIs}\})$  \\
&\underline{\textit{Selection rule for pre-selected variables}}\\
3 & $\mathfrak{u}_{\{10\}}(\{\texttt{Age},\texttt{Sex}, \texttt{Stroke},\texttt{Anemia}, \texttt{Malignancy}, \texttt{Liver diseases},$ \\
&\hspace{0.2cm} \texttt{History of major bleeding}, \texttt{Renal diseases}, \texttt{Antiplatelets}, \texttt{NSAIDs}\})\\
\hline
\end{tabular}\\
\caption{The selection rules for the running example.}\label{SRs}
    \end{center}
\end{table}
Rule 1.1 is needed since when \texttt{DOAC} is in the model, and if \texttt{High-dose-DOAC} is selected, then the interpretation of the coefficient of \texttt{High-dose-DOAC} is the contrast (e.g. log odds ratio) of high-dose-DOAC versus low-dose-DOAC, which is of interest. However, without \texttt{DOAC} in the model, these relevant interpretations would be lost. 
Rule 1.2 is needed since when \texttt{DOAC} is in the model, and if \texttt{Apixaban} is selected, then the interpretation of the coefficient of \texttt{Apixaban} is the contrast of Apixaban and Rivaroxaban. 
However, without \texttt{DOAC}, the coefficient of \texttt{Apixaban} would represent a contrast against warfarin and Rivaroxaban combined, which is less interpretable.  
The same rationale applies for \texttt{Dabigatran} in rule 1.3.
The rationale for rule 1.5 is that: 
1) according to strong heredity, both \texttt{High-dose-DOAC} and \texttt{High-adherence} must be in the model when the interaction of \texttt{High-dose-DOAC} and \texttt{High-adherence} is selected, 
2) when \texttt{High-dose-DOAC} is in the model, \texttt{DOAC} must be selected, which is justified by the rule 1.1, and 
3) the interaction of \texttt{High-dose-DOAC} and \texttt{High-adherence} represents a three-way interaction between \texttt{High-dose-DOAC}, \texttt{DOAC}, and \texttt{High-adherence}. Therefore, by the rationale of strong heredity, we must include the lower order interaction between \texttt{DOAC} and \texttt{High-adherence}. If we do not, we are assuming that \texttt{High-adherence} has the same impact whether a patient takes warfarin or low-dose-DOAC but the impact is different for a high-dose-DOAC. The coefficient of the main term \texttt{High-adherence} would then be less interpretable. The rules 1.4 and 2.1-2.4 are straight-forward applications of strong heredity. Rule 3 is based on the findings from \citet{lane2012use}.

We observed that all of the selection rules defined previously can be represented by unit rules and if-then rules. The unit rules applied here are special unit rules where the $\mathbb{C}$ in $\mathfrak{u}_{\mathbb{C}}(\mathbb{F})$ always equals $|\mathbb{F}|$. Thus we omit $\mathbb{C}$ in the following, and the selection dictionary of $\mathfrak{u}(\mathbb{F})$ is $\{\mathbb{F}\cup \mathbb{m}: \mathbb{m}\in \mathcal{P}(\mathbb{V})\setminus \mathbb{F}\}$. We next show, using the mappings between the above selection rules and selection dictionaries, the algorithms to derive the selection dictionary that respects all selection rules simultaneously using set operations on the rule-specific dictionaries in Algorithm \ref{algo1}. We also show the exhaustive search method in Algorithm \ref{algo2}. Algorithm \ref{algo1} constructs the selection dictionary by the unit dictionaries and their operations, whereas Algorithm \ref{algo2} eliminates the sets in $\mathcal{P}(\mathbb{V})$ that do not respect the rules. Since rule 3 forces the selection of variables, they have to be in each set of the selection dictionary. Both algorithms and the resulting selection dictionary are available in the Github \url{https://github.com/Guanbo-W/SelectionDictionary}.

After running the algorithms, we found that the cardinality of the selection dictionary corresponding to the combination of our 7 rules is 32512. Algorithm \ref{algo1} took 5 minutes while Algorithm \ref{algo2} took 3.5 seconds on a local computer without the use of parallel computing. When applying these algorithms to other applications, Algorithm \ref{algo2} can be more prone to errors, because it involves encoding each selection rule manually. Whereas Algorithm \ref{algo1} can be automated with generic functions. 


\begin{algorithm}
\caption{Steps of deriving the selection dictionary by set operations}
\label{algo1}
Denote the variables in rule 3 by $\mathbb{A}$. \\
Define $\mathbb{V}$ as a set that contains all candidate variables except $\mathbb{A}$.\\
Conduct steps 1 and 2 for each if-then rule (1.1-2.4.4) $\mathfrak{r}_j=\mathfrak{u}_{j,1}(\mathbb{F}_{j,1})\rightarrow\mathfrak{u}_{j,2}(\mathbb{F}_{j,2})$:\\
Step 1: Derive the unit dictionaries $\mathbb{D}_{\mathfrak{u}_{j,1}}=\{\mathbb{F}_{j,1}\cup \mathbb{m}: \mathbb{m}\in \mathcal{P}(\mathbb{V})\setminus \mathbb{F}_{j,1}\}$. Similarly for $\mathbb{D}_{\mathfrak{u}_{j,2}}$.\\
Step 2: Derive the selection dictionary of $\mathfrak{r}_{j}$ as $\mathbb{D}_{\mathfrak{r}_{j}}=\{\mathcal{P}(\mathbb{V})\setminus\mathbb{D}_{\mathfrak{u}_{j,1}}\}\cup(\mathbb{D}_{\mathfrak{u}_{j,1}}\cap\mathbb{D}_{\mathfrak{u}_{j,2}})$.\\
Step 3: Derive the selection dictionary that respects all rules 1.1-2.4.4 by $\mathbb{D}_{\mathfrak{r}}=\cap_{j}\mathbb{D}_{\mathfrak{r}_{j}}$.\\
Step 4: The final selection dictionary is $\{\mathbb{A}\cup\mathbb{n}: \mathbb{n}\in\mathbb{D}_{\mathfrak{r}}\}$. 
\end{algorithm}
\begin{algorithm}
\caption{Steps of deriving the selection dictionary by exhaustive search}
\label{algo2}
Denote the variables in rule 3 by $\mathbb{A}$. \\
Define $\mathbb{V}$ as a set containing all candidate variables except $\mathbb{A}$.\\
For each subset of $\mathcal{P}(\mathbb{V})$, denoted $\mathbb{x}$:\\
Step 1: for all if-then rules (1.1-2.4) denoted $\mathfrak{r}_j=\mathfrak{u}_{j,1}(\mathbb{F}_{j,1})\rightarrow\mathfrak{u}_{j,2}(\mathbb{F}_{j,2})$, check if $\mathbb{x}$ satisfies the following condition: 
($\mathbb{F}_{j,1}\in\mathbb{x}$ \texttt{and} $\mathbb{F}_{j,2}\in\mathbb{x}$) \texttt{or} $\mathbb{F}_{j,1}\notin\mathbb{x}$.\\
Step 2: Collect all the $\mathbb{x}$ that satisfy the above condition, and denote the collection as $\mathbb{D}_{\mathfrak{r}}$.\\
Step 3: The final selection dictionary is $\{\mathbb{A}\cup\mathbb{n}: \mathbb{n}\in\mathbb{D}_{\mathfrak{r}}\}$.
\end{algorithm}

\section{Deriving the grouping structure from the selection dictionary}\label{sec:SD-GS}
\subsection{For the LOGL}
Revisiting the ultimate goal, we aim to utilize a penalized regression to construct a coherent prediction model that incorporates pre-defined selection rules and dictionaries. In this section, we address the gap using LOGL specifically, due to its capacity to respect most types of selection rules. For the LOGL, we provide both theoretical justification and practical roadmaps for identifying grouping structures that respect the pre-defined selection rules. We then apply these results to our running example. 


We first review the LOGL. Denote the outcome and covariates by $Y$ and $\mathbb{V}$, respectively. Denote the coefficients of covariates in the penalized regression by $\boldsymbol{\beta}$. Define a set of latent coefficients $\bar{\boldsymbol{\alpha}}=(\boldsymbol{\alpha}^{\mathbb{g}_{i}})_{\mathbb{g}_{i}\in\mathbb{G}}$ such that $\sum_{i=1}^{I}\boldsymbol{\alpha}^{\mathbb{g}_{i}}=\boldsymbol{\beta}$, where $\boldsymbol{\alpha}^{\mathbb{g}_{i}}$ is a vector of the same length as $\boldsymbol{\beta}$ whose coordinates are non-zero for indices in the set $\mathbb{g}_i$ and 0 otherwise. A grouping structure $\mathbb{G}\coloneqq\{\mathbb{g}_{i},i=1,...,I\}$ is a set of non-empty subsets $\mathbb{g}_i\subseteq\mathbb{V}$ s.t.  $\cup_{i=1 }^{I}\mathbb{g}_{i}=\mathbb{V}$. 

Let $\ell(\boldsymbol{\beta})$ be a convex loss function, $\boldsymbol{\eta}$  a vector of hyperparameters, and  $\Omega(\boldsymbol{\boldsymbol{\alpha}}^{\mathbb{g}_{i}};\boldsymbol{\eta})$  the penalty term. The LOGL solves
$
    \min_{\boldsymbol{\beta}} \ell(\boldsymbol{\beta})+\sum_{\mathbb{g}_{i}\in\mathbb{G}}\omega_{\mathbb{g}_i}\Omega(\boldsymbol{\boldsymbol{\alpha}}^{\mathbb{g}_{i}};\boldsymbol{\eta}), 
$
where the level of penalization is controlled by the possibly multivariate hyper-parameter $\boldsymbol{\eta}$, and where the scalar $\omega_{\mathbb{g}_i}$ is a positive weight applied to the coefficients in the group $\mathbb{g}_i$.

In practice, the penalty function $\Omega(\cdot)$ can be an $L^{2}$ norm with single hyperparameter $\boldsymbol{\eta}=\lambda$ \citep{obozinski2011variable}, minimax concave penalty (MCP) with hyperparameters $\boldsymbol{\eta}=(\lambda,\gamma)$ \citep{zhang2010nearly,huang2012selective}, or smoothly clipped absolute deviation (SCAD) with $\boldsymbol{\eta}=(\lambda,\gamma)$ \citep{fan2001variable,breheny2015group}. The latter two penalties have the oracle property \citep{breheny2011coordinate} and retain the penalization rate of the $L^{2}$ norm for small coefficients, but continuously relax the rate of penalization as the absolute value of the coefficient increases. The rate of relaxation is larger in MCP, compared with SCAD. The specifications of $\Omega(\cdot)$ for these three penalties are given in web material 1, Section \ref{penalty}.

The LOGL operates as follows. The latent covariates (represented by their latent coefficients) specified in the same group are selected collectively. Once the algorithm is applied to the dataset, the variables in the selected groups are defined as the selected variables. Therefore, the selected variables are the union of the covariates in the groups not being selected. A notable aspect of the LOGL is its allowance for groups to overlap, contributing to its flexibility in accommodating a diverse array of selection rules.

Different specifications of grouping structures result in different combinations of subsets of variables that can be potentially selected. When the set of combinations under a given grouping structure is the same as the selection dictionary for a given selection rule, the grouping structure respects the selection rule. In the following theorem, we present the sufficient and necessary condition for a grouping structure to respect a selection rule when used with the LOGL.
\begin{theorem}
\label{th:ValidationThorem}
For a given $\mathbb{V}$, with LOGL, the sufficient and necessary condition of a grouping structure $\mathbb{G}\coloneqq\{\mathbb{g}_{i}, i=1,\dots,I\}$ to be congruent to a selection rule $\mathfrak{r}$ is $\mathbb{D}_{\mathfrak{r}}=\{\cup_{\mathbb{g}\in\mathbb{Q}_{j}}\mathbb{g}, j=1,\dots,2^{I}\}$, where $\mathbb{Q}_{j}, j=1,...,2^I$ are all unique subsets of $\mathbb{G}$.
\end{theorem}
\noindent 
The proof of Theorem~\ref{th:ValidationThorem} is given in web material 1, Section \ref{web materials_validation}.
Theorem \ref{th:ValidationThorem} provides us with an equation to check if LOGL can potentially respect a given selection rule, and if so, whether a specified grouping structure respects the selection rule. We show in web materials 1, Section \ref{GeneralPenalizedRegression} that such conditions can be generalized to the overlapping group Lasso \citep{wang2024structured} and general penalized regressions.

Even though Theorem \ref{th:ValidationThorem} does not tell us how to define a grouping structure for a selection rule, it gives us direct insight into how to postulate a grouping structure: starting from the selection rule dictionary, we seek all groups that satisfy the condition. We next give two examples of grouping structures with the LOGL.

\tikzstyle{arrow} = [thick,->,>=stealth]
\begin{figure}[h]
  \begin{subfigure}[]{.5\textwidth}
    \centering
      \begin{tikzpicture}[every node/.style={thick,inner sep=0pt}]
\node [] (a) at (-2.5,1){$A$};
\node [] (b1) at (-1,1){$B_{1}$};
\node [] (b2) at (0,1){$B_{2}$};
\node [] (ab1) at (-2,-1){$AB_{1}$};
\node [] (ab2) at (-1,-1){$AB_{2}$};
\draw [arrow] (a) -- (ab1);
\draw [arrow] (a) -- (ab2);
\draw [arrow] (b1) -- (ab1);
\draw [arrow] (b2) -- (ab2);
\draw [red] (-1.255,0.2) ellipse [x radius = 2.2cm, y radius = 1.7cm];
\draw [blue] (-2.5,1) ellipse [x radius = 0.3cm, y radius = 0.3cm];
\draw [blue] (-0.5,1) ellipse [x radius = 1cm, y radius = 0.3cm];
\end{tikzpicture}
  \end{subfigure}
  \begin{subfigure}[]{.5\textwidth}
    \centering
      \begin{tikzpicture}[every node/.style={thick,inner sep=0pt}]
\node [] (a) at (-2.5,1){$A$};
\node [] (b1) at (-1,1){$B_{1}$};
\node [] (b2) at (0,1){$B_{2}$};
\node [] (ab1) at (-2,-1){$AB_{1}$};
\node [] (ab2) at (-1,-1){$AB_{2}$};
\draw [arrow] (a) -- (ab1);
\draw [arrow] (a) -- (ab2);
\draw [arrow] (b1) -- (ab1);
\draw [arrow] (b2) -- (ab2);
\draw [orange] [rotate=-55] (-1.1,-1.6) ellipse [x radius = 1.8cm, y radius = 0.9cm];
\draw [green] [rotate=50] (-0.5,0.8) ellipse [x radius = 2cm, y radius = 1.1cm];
\draw [blue] (-2.5,1) ellipse [x radius = 0.3cm, y radius = 0.3cm];
\draw [blue] (-0.5,1) ellipse [x radius = 1cm, y radius = 0.3cm];
\end{tikzpicture}
  \end{subfigure}
\caption{The left and right panels correspond to the grouping structures in examples 1 and 2, respectively.}\label{Fig:StructureExamples}
\end{figure}
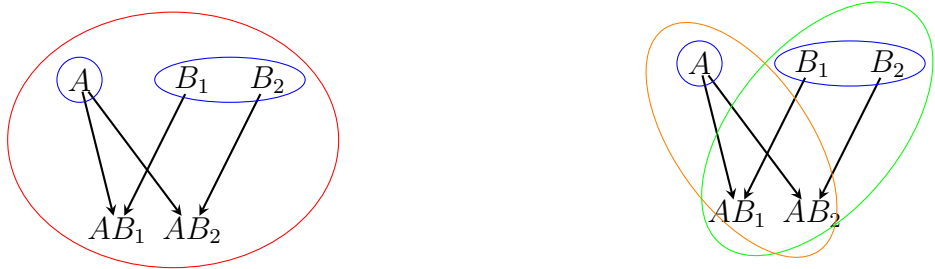
Suppose $\mathbb{V}=\{A, B_{1}, B_{2}, AB_{1}, AB_{2}\}$, where $A$ is a binary variable, $B_1$ and $B_2$ are indicators representing two levels of a three-level categorical variable $B$, and $AB_{1}, AB_{2}$ are terms representing the interaction between $A$ and $B$. We are interested in incorporating the selection rules ``if the interaction terms are selected, then the main terms must be selected'' (strong heredity), and ``if the interactions are selected, then at least one of the main terms must be selected'' (weak heredity). To express the selection rules, we define $\mathbb{F}_1=\{AB_{1}, AB_{2}\}$, $\mathbb{F}_2=\{B_{1}, B_{2}\}$ and $\mathbb{F}_3=\{A, B_{1}, B_{2}\}$. Examples 1 and 2 express the selection rules using our notation, and the left- and right-hand panels of Figure \ref{Fig:StructureExamples} give the corresponding grouping structures. We can see that some groups are overlapped.\\
\textbf{Example 1} LOGL can achieve categorical interaction selection with strong heredity \citep{lim2015learning}, $\big(\mathfrak{u}_{\{0, 2\}}(\mathbb{F}_1)\land \mathfrak{u}_{\{0,2\}}(\mathbb{F}_2)\big)\land\big(
\mathfrak{u}_{\{1, 2\}}(\mathbb{F}_1)\rightarrow
\mathfrak{u}_{\{3\}}(\mathbb{F}_3)\big)$. The corresponding grouping structures are represented by the left-hand panel of Figure \ref{Fig:StructureExamples}. \\
\textbf{Example 2} LOGL can also achieve categorical interaction selection with weak heredity, $\big(\mathfrak{u}_{\{0, 2\}}(\mathbb{F}_1)\land \mathfrak{u}_{\{0,2\}}(\mathbb{F}_2)\}\big)  \land\big(
\mathfrak{u}_{\{1, 2\}}(\mathbb{F}_1)\rightarrow
\mathfrak{u}_{\{1,3\}}(\mathbb{F}_3)\big)$. The corresponding grouping structure is represented in the right-hand panel of Figure \ref{Fig:StructureExamples}.
 
For the above selection rules, it may be possible to guess and verify the corresponding grouping structures. However, it remains a non-trivial task to actually construct the grouping structures for complex selection rules. We next develop roadmaps for generic grouping structure identification for some common selection rules, including the ones seen in this application.

Define $\mathbb{A}=\{A_1,\dots\mathrel{,}A_n\}$, and $\mathbb{B}=\{B_1,\dots,B_m\}$ as two non-overlapping sets of binary or continuous variables. Table \ref{tab:roadmap} gives four types of selection rules with the corresponding roadmaps for constructing the corresponding grouping structure. As an implication of Theorem \ref{th:ValidationThorem}, when combining those rules, the resulting grouping structure of any combined selection rules in Table \ref{tab:roadmap} is the union of the grouping structures of each selection rule. Some caveats for using the roadmaps are given in web material 1, Section \ref{App:roadmap}. With such a roadmap, we can cover most of the selection rules that can be incorporated by the LOGL, so that with the pre-defined selection rules, we can directly draw the grouping structures.
\begin{table}[]
    \centering
    \begin{tabular}{p{7cm}p{8cm}}\hline
    \textbf{Selection rule} & \textbf{Roadmap of grouping variables} \\\hline\hline
     If all variables in $\mathbb{A}$ are selected, then all variables in $\mathbb{B}$ must be selected& 
    Specify $m$ single-variable groups for $B_1\dots,B_m$, then specify another group that contains $\mathbb{A}$ and $\mathbb{B}$.\\ 
     If at least one variable in $\mathbb{A}$ is selected, then all variables in $\mathbb{B}$ must be selected& 
    Specify $m$ single-variable groups for $B_1\dots,B_m$, then specify $n$  groups, each  containing $\mathbb{B}$ and $A_i$, for $i=1,\dots,n$.\\ 
     If all variables in $\mathbb{A}$ are selected, then at least one variable in $\mathbb{B}$ must be selected & 
    Specify $m$ single-variable groups for $B_1\dots,B_m$, then specify $m$ groups, each containing $B_j$ and $\mathbb{A}$ for $j=1,\dots,m$.\\
     If at least one variable in $\mathbb{A}$ is selected, then at least one variable in $\mathbb{B}$ must be selected & 
    Specify $m$ single-variable groups for $B_1\dots,B_m$, then specify $n\times m$ groups, where each group contains one variable from each set, i.e. each group contains $A_i$ and $B_j$ for some $i=1,\dots,n, j=1,\dots,m$.\\\hline
    \end{tabular}
    \caption{Roadmaps of grouping structure identification for the LOGL; $\mathbb{A}=\{A_1,\dots\mathrel{,}A_n\}$, and $\mathbb{B}=\{B_1,\dots,B_m\}$ are two non-overlapping sets of binary or continuous variables.}
    \label{tab:roadmap}
\end{table}
\noindent 

\subsection{Application}
Finally, we use the running motivating example to illustrate how to use Theorem \ref{th:ValidationThorem} and the roadmap to identify the grouping structure used with LOGL to incorporate the selection rules defined in Table \ref{SRs}. Since our selection rules 1.1-2.4 are represented in the form of ``if one variable is selected, then some other variables must be selected'', we will only need the first roadmap in Table \ref{tab:roadmap}. For example, for rule 1.4, we create single-variable groups for \texttt{DOAC} and \texttt{High-adherence} and a third group that contains \texttt{DOAC}, \texttt{High-adherence}, and their interaction. We do this for each rule 1.1-2.4, removing duplicate groups.
To implement rule 3, which cannot strictly be respected by LOGL, we modified it to ``select either none or all of the 10 variables'', which is practically the same as the original version (details in web material 1, Section  \ref{App:rule3}).  The modified rule can be implemented in the LOGL by creating a group for those 10 variables. 

Following the above steps, we established the grouping structure in the application, which is exhaustively listed in web material 1, Section \ref{GS_app}. Then we used the \texttt{R} package \texttt{grpregOverlap} \citep{RgrpregOverlap} to implement the LOGL with various penalties. The weights for each group were set to be the square root of the number of variables in the group \citep{obozinski2011variable}. We next compare the results from the LOGL with those of lasso \citep{tibshirani1996regression} and adaptive Lasso \citep{zou2006adaptive}. Note that unlike Lasso, adaptive Lasso possesses the oracle property, namely, for large sample size, it performs as well as if the true underlying model were given in advance and thus the results are more trustworthy \citep{zou2006adaptive}. All code for this analysis is available at \url{https://github.com/Guanbo-W/SelectionDictionary}.
\section{Results}
\label{results}
We present the results of analyzing the data using the techniques developed above to incorporate the selection rules defined in Section \ref{sec:CompoundSelectionRule}. 

The overall rate of the outcome (major bleeding) is 3.47 per 100 person-years, and the percentage of patients who experienced major bleeding is 3.1\%. The means and standard errors or proportions of all covariate data stratified by the outcome are given in web material 1, Section \ref{appendix:TableOneVariables}. The summary statistics of variables stratified by high-dose-DOAC, low-dose-DOAC and warfarin are given in web material 1, Section \ref{SSSD}. Crude (univariate analyses) and adjusted odds ratios (obtained from the logistic regression model adjusting for key covariates) and 95\% confidence intervals are given in web material 1, Section \ref{Crud_Adjusted}.

Table \ref{TableOneVariables_res} gives the odds ratios for each variable estimated by Lasso, Adaptive Lasso (ALasso), and LOGL, with the latter under different penalties. For all methods, we selected the tuning parameter $\lambda$ at the minimum cross-validated risk, the tuning parameter $\gamma$ for MCP and SCAD is set to 3 and 4 respectively. 

\footnotesize
\begin{longtable}{p{9cm}lllll}
\hline
 & \multicolumn{2}{c}{\textbf{Non-grouped}} & \multicolumn{3}{c}{\textbf{LOGL}*}\\ \cmidrule(lr){2-3}\cmidrule(lr){4-6}
\textbf{Variable name} & Lasso & ALasso** & $L^{2}$ & MCP & SCAD \\
\hline\hline
cross-validated risk &0.271 & 0.060& 0.031 & 0.031 & 0.031\\
\multicolumn{6}{c}{\textbf{Baseline covariates}}\\
1. Age ($\geqslant$75/$<$75) &1.21& 1.09&1.24 & 1.24 & 1.24\\
2. Sex (female/male) & 0.88&- &0.87 & 0.86&0.86\\
3. CHA$_{2}$DS$_{2}$-VASc Score ($\geqslant$3/$<$3) &1.19 &1.04 &- &1.24 &1.24\\
\multicolumn{6}{c}{\textbf{Comorbidities within 3 years before cohort entry}}\\
4. Stroke (yes/no)&0.95 & -& 0.95 &0.94 &0.94\\
5. Anemia (yes/no) &1.35 &1.35& 1.35 &1.38 &1.38\\
6. Malignancy (yes/no) & 1.07 &-&1.08 &1.09 &1.09\\
7. Liver disease (yes/no) & 1.92&1.98&  1.92 & 2.01&2.01\\
8. History of major bleeding (yes/no)&1.59 &1.60& 1.53 & 1.62&1.62\\
9. Renal diseases (yes/no)&- &-& 0.97&0.96 &0.96\\
10. Heart disease (yes/no) &1.20 &1.13& 1.21 & 1.22&1.22\\
11. Diabetes (yes/no) &1.28 & 1.22& 1.28 & 1.31&1.31\\
12. COPD/asthma (yes/no)&1.13 &1.02& 1.13 &1.15 &1.15\\
13. Dyslipidemia (yes/no)&1.07 &-&  1.07 &1.09 &1.09\\
\multicolumn{6}{c}{\textbf{OAC use at cohort entry}}\\
14. DOAC (DOACs/warfarin)  &1.18 &-& 1.29 & 1.39&1.39\\
15. Apixaban (yes/no), ref: Rivaroxaban& 0.69& 0.72& 0.67& 0.62&0.62\\
16. Dabigatran (yes/no), ref: Rivaroxaban& 1.11 & -& 1.03& -&-\\
17. High-dose-DOAC (high-dose-DOACs/low-dose-DOACs or wafarin) & 0.88 &0.89& 0.81 & 0.81&0.81\\
18. High-adherence (high-adherence/low-adherence) &-&-& 1.01  & -&-\\
19. Interaction of DOAC and High-adherence&-& -&0.96& -&-\\
20. Interaction of High-dose-DOAC and High-adherence & -&-&1.05 & -&-\\
\multicolumn{6}{c}{\textbf{Concomitant medication use within 2 weeks before cohort entry}}\\
21. Antiplatelets (yes/no) &1.35 &1.17& 1.30 &1.51 &1.51\\
22. NSAIDs (yes/no)&- &-& 1.28 & 1.35&1.35\\
23. Antidepressants (yes/no)& 1.13 &-& 1.15 & 1.16&1.16\\
24. PPIs (yes/no)& 0.87 &-&0.88 &0.80 &0.80\\
\multicolumn{6}{c}{\textbf{Potential drug-drug interaction}}\\
25. Interaction of DOAC and Antiplatelets &0.81 &-& 0.86 & 0.68&0.68\\
26. Interaction of DOAC and NSAIDs &1.41 &1.01& - & -&-\\
27. Interaction of DOAC and Antidepressants &-&-& 0.94 & -&-\\
28. Interaction of DOAC and PPIs &0.91 &-& 0.92 & -&-\\
\hline

\caption{Coefficients estimates from various methods. - indicates the variable was not selected. *LOGL: latent overlapping group Lasso. **ALasso: Adaptive Lasso.}
\label{TableOneVariables_res}%
\end{longtable}
\normalsize
\noindent
From the results, we see that Lasso selected the most variables, while adaptive Lasso selected the fewest variables. The resulting (averaged 10-fold) cross-validated risks were 0.271 and 0.060, respectively. However, neither method can incorporate selection rules. In fact, we see that both of them violated rule 3. Both Lasso and adaptive Lasso selected the interaction of \texttt{DOAC} and \texttt{NSAIDs}, but not \texttt{NSAIDs}, violating strong heredity. Adaptive Lasso selected \texttt{Dabigatran}, but not \texttt{DOAC}, which complicates the interpretation of the coefficient of \texttt{Dabigatran}. In addition, adaptive Lasso selected \texttt{Dose} but not \texttt{DOAC}, which gives us a contrast between high-dose-DOACs versus low-dose-DOACs or warfarin, which is less interpretable than a contrast between high and low-dose-DOACs. 

The LOGL reduced the cross-validated risk to 0.031. By design, this method respected the selection rules, resulting in a model with better interpretability. Though the cross-validated risks of using different penalties in the LOGL were the same, the set of variables selected were different depending on the penalty type. The results derived from MCP and SCAD were similar, likely due to both of them having oracle properties. These methods penalize coefficients less when the estimated odds ratios strongly deviate from 1. So we see that the estimated odds ratios from MCP and SCAD were further from 1 compared to those using the $L^{2}$ penalty. Additionally, these two penalties resulted in fewer variables being selected. In contrast with the $L^{2}$ penalty, they did not select \texttt{High-adherence}, the interaction of \texttt{DOAC} and \texttt{High-adherence}, the interaction of \texttt{High-adherence} and \texttt{High-dose-DOAC}, the interaction of \texttt{DOAC} and \texttt{Antiplatelets}, and the interaction of \texttt{DOAC} and \texttt{PPIs}, though the estimated odds ratios of these variables under the $L^{2}$ penalty were close to 1.

From MCP and SCAD, the most predictive factors associated with higher risk of major bleeding (odds ratios above 1.50) were \texttt{liver disease} (2.01), \texttt{History of major bleeding} (1.62) and \texttt{Antiplatelets} (1.51). Important predictors that were associated with lower risk of major bleeding (odds ratios below 0.8) are \texttt{Dabigatran} (0.62) and the interaction of \texttt{DOAC} and \texttt{Antiplatelets} (0.68). From the results, we can also summarize the estimated odds ratios of taking different types of DOACs versus warfarin in Table \ref{tab: ratioanle}, which are also of interest. For instance, the estimated odds ratio of \texttt{High-dose-Apixaban} versus \texttt{warfarin} is 0.7. The method to obtain these additional contrasts is given in web material 1, Section \ref{rationale}.
\begin{table}[h]
    \centering
    \begin{tabular}{cc}
    \hline
    Contrasts with warfarin as reference   &   Estimated odds ratios\\
    \hline\hline
    High-dose-Apixaban   & 0.70\\
    High-dose-Dabigatran     & 1.13\\
    High-dose-Rivaroxaban    & 1.13 \\
    Low-dose-Apixaban    & 0.86\\
    Low-dose-Dabigatran     & 1.39\\
    Low-dose-Rivaroxaban   & 1.39\\
   \hline
\hline
    \end{tabular}
    \caption{Estimated odds ratios of taking different types of DOACs versus warfarin from the selected model by the LOGL MCP/SCAD}
    \label{tab: ratioanle}
\end{table}
\section{Discussion}
\label{discussion}
In this work, we developed practical algorithms for deriving selection dictionaries from selection rules and provided guidance on how to effectively group variables used with the LOGL to respect the pre-specified selection rules. Together with the work by \citet{wang2023general}, our aim is to equip practitioners with both theoretical understanding and practical methods for constructing coherent prediction models that incorporate complex selection rules. 
This work also represents the first comprehensive implementation of the framework of structured variable selection \citep{wang2023general}, which we used to define a complex series of selection rules in a health application to identify predictors of major bleeding among hospitalized hypertensive patients using OACs for atrial fibrillation. We then followed our roadmaps to integrate the selection rules into LOGL in order to obtain a more interpretable prediction model. 

The proposed methods can be easily implemented if the set of covariates that are affected by the selection rules is low-dimensional or can be divided into non-overlapping low-dimensional subsets, regardless of the dimension of the total number of variables. In such scenarios, one does not need to derive the selection dictionary for all variables. For example, suppose that only 10 variables are constrained by selection rules, and the other 1000 variables can be independently selected. We can first derive the sub-selection dictionary regarding these 10 variables (treating $\mathbb{V}$ as the set of these 10 variables), and then identify the sub-grouping structure for the 10 variables. Then the remainder of the complete grouping structure involves an additional 1000 single-variable groups for each of the remaining variables.

Even though the LOGL is the most versatile variable selection technique, there are some selection rules that it cannot respect. Examples include ``select a number (between 0 and the cardinality of the subset) of variables in a subset of candidate variables'', and ``if a number of variables in a subset of candidate variables is selected, then select a number of variables in a (possibly distinct) subset of candidate variables''. Given these limitations of the LOGL, future work could focus on developing more general regularization methods that can respect an arbitrary selection rule. 
Another limitation of the application of the LOGL is that post-selection inference has not yet been developed for this method, so that post-selection confidence intervals are not currently available. 


Coherent prediction modeling can guide clinicians in identifying patients at risk of an outcome by highlighting which factors predict the outcome of interest and their importance in prediction. However, when the unique goal of modeling is the determination of patient risk, black box methods (i.e. those not restricted to a semiparametric model to the extent that the fitted model cannot be interpreted by the ultimate user) may be more accurate. They may thus be preferred as a decision-making aid. However, much clinical practice remains unassisted by such algorithms, and because the acceptance of a new prediction model may depend on face validity, interpretable and coherent modeling continues to provide important contributions to medical knowledge.  
\section*{Web materials}
Web material 1 contains some technical details of the method and numerical results from the analysis.\\
Web material 2 is the variable definitions of variables considered in the analysis. (available upon request)\\ 
Codes for deriving the selection dictionary, and analysis, and the resulting selection dictionary are seen in the Github \url{https://github.com/Guanbo-W/SelectionDictionary}.
\section*{Data availability statement}
The data analyzed in this work are not allowed to share due to the confidentiality.
\section*{Funding}
The study was supported by the Heart and Stroke Foundation of Canada (G-17-0018326) and the Réseau Québécois de Recherche sur les Médicaments (RQRM). Please refer to \url{https://www.heartandstroke.ca/} and \url{http://www.frqs.gouv.qc.ca/en/}. G Wang is supported by a Fonds de Recherche du Québec—Santé Doctoral Training Award (272161) and the Faculté de Pharmacie, Université de Montréal. ME Schnitzer is supported by a Canadian Institutes of Health Research Canada Research Chair tier 2 and a Natural Sciences and Engineering Research Council of Canada Discovery Grant. RW Platt is supported by a CIHR Foundation Grant (FDN-143297). 
\section*{Acknowledgements}
We would like to thank the RAMQ and Quebec Health Ministry for providing assistance in handling the data and the Commission d’accès à l’information for authorizing the study.
\bibliography{0refs}%
\clearpage

\appendix
\noindent
{\Large \textbf{Web material 1 of ``Integrating complex selection rules into the latent overlapping group Lasso for constructing coherent prediction models'' \\
}}
\clearpage
\section{Population-based cohort definition flowchart}
\label{exclusion}
\begin{figure}[H]
    \centering
    \includegraphics[scale=0.26]{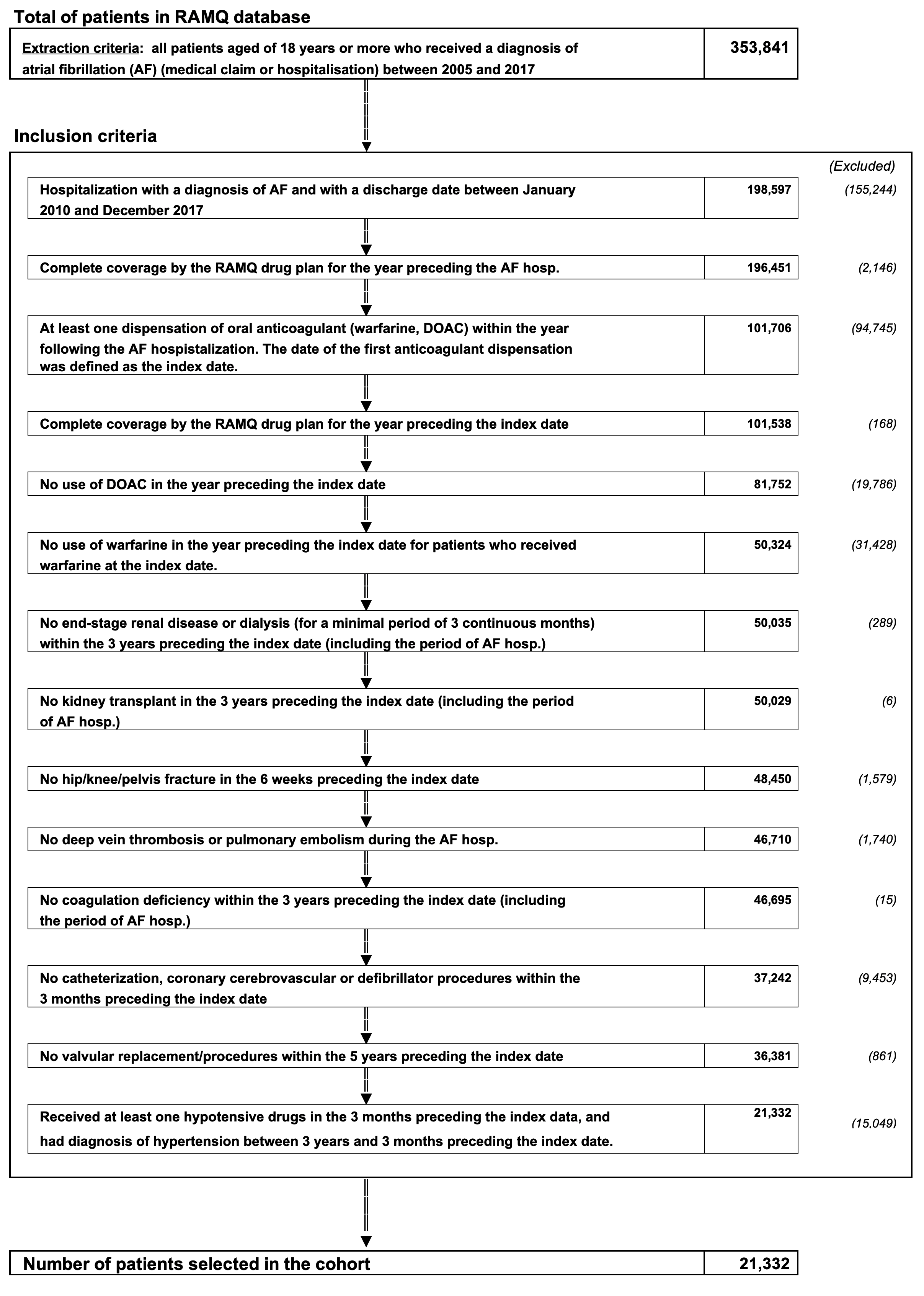}\\
    \caption{Population-based cohort definition flowchart.}
    \label{fig:exclusion}
\end{figure}
\section{Correspondence between all possible drugs taken and variable coding}\label{app:corr}
See Table \ref{tab:corr} for the correspondence between all possible drugs taken and variable coding.
\begin{table}[]
    \centering
    \begin{tabular}{ccccc}\hline
       Drug taken  & \texttt{DOAC} & \texttt{Apixaban}&\texttt{Dabigatran}&\texttt{High-dose-DOAC} \\
       \hline\hline
      warfarin   & 0 & 0 & 0 & 0\\
      High-dose-Apixaban   & 1& 1& 0& 1\\
    High-dose-Dabigatran     & 1& 0& 1& 1\\
    High-dose-Rivaroxaban    &1 &0 &0 & 1 \\
    Low-dose-Apixaban    & 1& 1& 0&0 \\
    Low-dose-Dabigatran     &1 &0 &1 &0 \\
    Low-dose-Rivaroxaban   & 1&0 &0 & 0\\
    \hline
    \end{tabular}
    \caption{Correspondence between all possible drugs taken and variable coding}
    \label{tab:corr}
\end{table}

\section{Different specification of penalties}
\label{penalty}
Table \ref{Different specification of penalties} shows the different specification of the penalties.
\begin{table}[ht]
    \centering
    \begin{tabular}{LL}
    \hline
    \textbf{Penalty}     &   \textbf{Specification} \\\hline\hline
    L^{2}     & 
    \Omega(\boldsymbol{x};\boldsymbol{\lambda})=\lambda\norm{\boldsymbol{x}}_{2}
   \\\\
    \text{MCP} & 
    \Omega(\boldsymbol{x};\boldsymbol{\lambda})=
    \begin{cases}
    \lambda|\boldsymbol{x}|-\frac{\boldsymbol{x}^{2}}{2\gamma}, &\text{if } \lvert \boldsymbol{x} \rvert \leqslant\gamma\lambda\\
    \frac{1}{2} \gamma\lambda^{2}, &\text{if } \lvert \boldsymbol{x} \rvert>\gamma\lambda
    \end{cases} \text{ } \gamma>1
   \\\\
    \text{SCAD} & 
    \Omega(\boldsymbol{x};\boldsymbol{\lambda})=
    \begin{cases}
    \lambda|\boldsymbol{x}|, & \text{if } \lvert \boldsymbol{x} \rvert \leqslant\lambda,\\
    \frac{2\gamma\lambda|\boldsymbol{x}|-\boldsymbol{x}^{2}-\lambda^{2}}{2(\gamma-1)}, & \text{if } \lambda<\lvert \boldsymbol{x} \rvert<\gamma\lambda,\\
    \frac{\lambda^{2}(\gamma+1)}{2}, &\text{if }  \lvert \boldsymbol{x} \rvert\geqslant\gamma\lambda
    \end{cases} \text{ } \gamma>2
    \\\hline
    \end{tabular}
    \caption{Different specification of penalties. In MCP and SCAD, we set $\boldsymbol{x}=\norm{\boldsymbol{\alpha}^{\mathbb{g}_{i}}}_{2}$.}
    \label{Different specification of penalties}
\end{table}
\section{Proof of Theorem \ref{th:ValidationThorem}}
\label{web materials_validation}
\begin{proof}
The variables being selected by LOGL is the union of groups of variables whose latent coefficients are estimated as non-zero. Mathematically, let $\text{supp}(\hat{\mathbb{V}})$ be the variables being selected by LOGL, and let $\mathbb{Q}\subseteq\mathbb{G}$ be the set of groups $\mathbb{g}$ with the estimate of latent coefficients $\hat{\boldsymbol{\alpha}}_{|\mathbb{g}}$ such that   $\hat{\boldsymbol{\alpha}}_{|\mathbb{g}}\neq\mathbf{0}$. Then $\text{supp}(\hat{\mathbb{V}})=\cup_{\mathbb{g}\in\mathbb{Q}}\mathbb{g}$. Note that $\mathbb{Q}$ is an element of the power set of $\mathbb{G}$.\\
For a given $\mathbb{V}$ with grouping structure $\mathbb{G}\coloneqq\{\mathbb{g}_{i},i=1,\dots,I\}$, by definition of grouping structure $\cup_{i=1}^{I}\mathbb{g}_{i}=\mathbb{V}$. Depending on the estimation, there are $2^I$ possible combinations of groups with non-zero latent coefficients $\hat{\boldsymbol{\alpha}}_{|\mathbb{g}}$. That is, there are $2^I$ possible $\mathbb{Q}$s. List these as $\mathbb{Q}_j, j=1,...,2^I$.  Given $\mathbb{Q}_j$, the variables that are selected into the model by LOGL are the covariates in [any of the groups in] $\mathbb{Q}_j$. We can represent these covariates by $\mathbb{d}_{j}=\cup_{\mathbb{g}\in\mathbb{Q}_{j}}\mathbb{g}$. By definition, each $\mathbb{d}_j,j=1,...,2^I$ is an element of the dictionary, i.e. one possible set of covariates that can be selected. So the congruent dictionary under this penalization structure $(\mathcal{M},\mathbb{G})$ is $\mathbb{D}_{\mathbb{G}}^{\mathcal{M}}=\{\mathbb{d}_j,j=1,...,2^I\}$. So any selection rule with congruent dictionary equal to $\mathbb{D}_{\mathbb{G}}^{\mathcal{M}}$  is congruent to the penalization structure $(\mathcal{M},\mathbb{G})$. Every step is necessary and sufficient so this completes the proof.

\end{proof}

\section{Generalizing Theorem~\ref{th:ValidationThorem} to the overlapping group Lasso and general penalized regressions}\label{GeneralPenalizedRegression}
The theorem below shows the condition of a grouping structure of the overlapping group Lasso being congruent to a selection rule
\begin{theorem}
For a given $\mathbb{V}$, with overlapping group Lasso, the necessary condition of a grouping structure $\mathbb{G}\coloneqq\{\mathbb{g}_{i}, i=1,\dots,I\}$ being congruent to a selection rule $\mathfrak{r}$ is $\mathbb{D}_{\mathfrak{r}}\setminus\mathbb{V}=\{(\cup_{\mathbb{g}\in\mathbb{Q}_{j}}\mathbb{g})^{\mathsf{c}}, j=1,\dots,2^{I}\}$, where $\mathbb{Q}_{j},j=1,...,2^I$ are all unique subsets of $\mathbb{G}$.
\end{theorem}
To develop the sufficient and necessary condition for general penalized regression, we first review the general penalized regressions, and define some new mathematical objects.

Many existing penalized regression methods (i.e. regularization methods) can respect non-trivial selection rules \citep{yuan2006model,campbell2017within,simon2013sparse,breheny2009penalized,Haris_2016,yuan2011efficient,jenatton2011structured,obozinski2011variable}. In fact, the different regularization methods were developed in order to respect different types of rules. In this section, we formalize the framework for penalization structures and describe how it connects to the selection rules and dictionaries.

For a given covariate set $\mathbb{V}$ and outcome $Y$, let $D=(\mathbb{V},Y)$ and denote the coefficients of covariates in the penalized regression by $\boldsymbol{\beta}$. Suppose the data are centered at 0, so that we omit the intercept, the penalized regression solves
$$
    \min_{\boldsymbol{\beta}} \ell(\boldsymbol{\beta})+\Omega(\boldsymbol{\beta},\boldsymbol{\theta}),
$$
where $\ell(\boldsymbol{\beta})$ is a convex loss function, $\boldsymbol{\theta}$ is a vector of hyper parameters, and penalty term $\Omega(\boldsymbol{\beta},\boldsymbol{\theta})$ is a function of $\boldsymbol{\beta}$ and $\boldsymbol{\theta}$. Different specifications of $\Omega$ result in different regularization methods, with different variable selection results. 
A grouping structure $\mathbb{G}\coloneqq\{\mathbb{g}_{i},i=1,...,I\}$ is a set of   non-empty subsets $\mathbb{g}_i\subseteq\mathbb{V}$ s.t.  $\cup_{i=1 }^{I}\mathbb{g}_{i}=\mathbb{V}$. Let $\boldsymbol{\beta}_{|\mathbb{g}}$ be a vector of the same length as $\boldsymbol{\beta}$ whose coordinates are equal to those of $\boldsymbol{\beta}$ for indices in the set $\mathbb{g}$ and 0 otherwise. Notation $\norm{\cdot}_{q}$ indicates the $L^{q}$ norm.

Table \ref{tab:penalties} summarizes five types of penalties with some key rules that they can respect. Each regularization method $\mathcal{M}$ has restrictions on which grouping structures are allowed. We say that the grouping structure is not compatible with a regularization method when the grouping structure does not satisfy the method's restrictions.  For example, when two groups overlap (i.e., include the same variables), the grouping structure $\mathbb{G}$ is not compatible with group Lasso.
\begin{table}[htbp]
    \begin{center}
    \begin{tabular}{lp{6cm}lp{5.9cm}}
    \hline
    $\mathcal{M}$     &  $\Omega(\boldsymbol{\beta};\boldsymbol{\theta})$& \textbf{Condition} & \textbf{Key rules}\\\hline\hline
    Lasso     &  $\lambda\sum_{\mathbb{g}\in\mathbb{G}}\norm{\boldsymbol{\beta}_{|\mathbb{g}}}_{1}$ & $\mathbb{g}=\mathbb{V}$& $\mathfrak{u}_{\{0,\dots,|\mathbb{V}|\}}(\mathbb{V})$\\
    GL     &  $\lambda\sum_{\mathbb{g}\in\mathbb{G}}\omega_{\mathbb{g}}\norm{\boldsymbol{\beta}_{|\mathbb{g}}}_{2}$ & $\mathbb{g}_{i}\cap\mathbb{g}_{j}=\emptyset$& $\land_{i}\mathfrak{u}_{\{0,|\mathbb{g}_{i}|\}}(\mathbb{g}_{i})$\\
    EGL    &  $\lambda\sum_{\mathbb{g}\in\mathbb{G}}\norm{\boldsymbol{\beta}_{|\mathbb{g}}}_{1}^{2}$ & $\mathbb{g}_{i}\cap\mathbb{g}_{j}=\emptyset$& $\land_{i}\mathfrak{u}_{\{1,\dots,|\mathbb{g}_{i}|\}}(\mathbb{g}_{i})$\\
    SGL & $(1-\gamma)\lambda\sum_{\mathbb{g}\in\mathbb{G}}\omega_{\mathbb{g}}\norm{\boldsymbol{\beta}_{|\mathbb{g}}}_{2}+\gamma\lambda\norm{\boldsymbol{\beta}}_{1}$ & $\mathbb{g}_{i}\cap\mathbb{g}_{j}=\emptyset$& $\land_{i}\mathfrak{u}_{\{0, \dots, |\mathbb{g}_{i}|\}}(\mathbb{g}_{i})\Rightarrow\land_{i}\mathfrak{u}_{\{1, \dots, |\mathbb{g}_{i}|\}}(\mathbb{g}_{i})$\\
    LOGL    &  $\lambda\sum_{\mathbb{g}\in\mathbb{G}}\omega_{\mathbb{g}}\norm{\boldsymbol{\alpha}_{|\mathbb{g}}}_{2}^{*}$, $\sum_{\mathbb{g}\in\mathbb{G}}\boldsymbol{\alpha}_{|\mathbb{g}}=\boldsymbol{\beta}$& NA & $\land_{i,j}\big[\mathfrak{u}_{\mathbb{C}_{i}}(\mathbb{g}_{i})\rightarrow\mathfrak{u}_{\mathbb{C}_{j}}(\mathbb{g}_{j})\big]$\\
    \hline
    \end{tabular}
    \end{center}
    \scriptsize{
GL: Group Lasso, EGL: Exclusive group Lasso, SGL: Sparse group Lasso, LOGL: latent overlapping group Lasso. }
    \caption{Summary of some penalization methods}
    \label{tab:penalties}
\end{table}
\noindent
Given a selection rule and some existing regularization method that could potentially satisfy the selection rule, our framework aims to illustrate how to specify $\boldsymbol{\beta}_{|\mathbb{g}}$, i.e., how to group variables in the penalty term $\Omega$. The framework can be generalized to accommodate more complicated penalties.

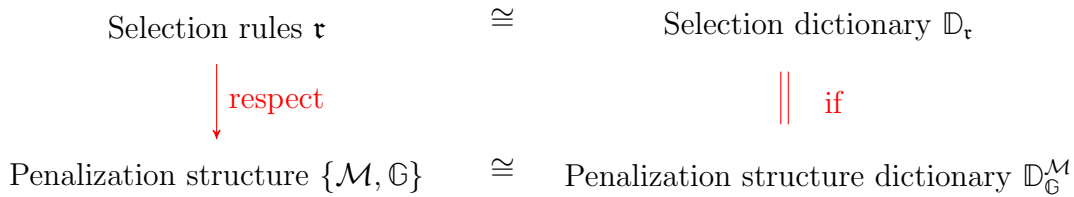
\begin{figure}[h]
    \centering
\begin{tikzpicture}
    \node[] (1) at (8,2) {Selection dictionary $\mathbb{D}_{\mathfrak{r}}$};
    \node[] (2) at (8,0) {Penalization structure dictionary $\mathbb{D}_{\mathbb{G}}^{\mathcal{M}}$};
    \node[] (3) at (0,2) {Selection rules $\mathfrak{r}$};
    \node[] (4) at (0,0) {Penalization structure $\{\mathcal{M},\mathbb{G}\}$};
    \node[] (5) at (8.2,1) {$\color{red}\text{if}$};
    \node[] (6) at (0.8,1) {$\color{red}\text{respect}$};
    \node[] (7) at (3.8,2.1) {$\cong$};
    \node[] (8) at (3.8,0.1) {$\cong$};
     \draw[red,->]        (0,1.5)   -- (0,0.5);
     \draw[red, -]        (7.5,1.4)   -- (7.5,0.7);
     \draw[red, -]        (7.6,1.4)   -- (7.6,0.7);
\end{tikzpicture}
    \caption{Conceptual structure of defined objects}
    \label{fig:Concept}
\end{figure}

We first characterize a ``penalization structure'' according to a given method and grouping of variables in the penalty term.
\begin{definition}[Grouping structure and penalization structure]
For a given $\mathbb{V}$, a \textbf{penalization structure} consists of a \textbf{grouping structure} and a compatible regularization method $\mathcal{M}$. A grouping structure  $\mathbb{G}\coloneqq\{\mathbb{g}_{i}\}$ s.t. $\cup_{i=1}^{I}\mathbb{g}_{i}=\mathbb{V}$ is any collection of non-empty subsets of $\mathbb{V}$, whose union is $\mathbb{V}$. 
\end{definition}
\noindent
A grouping structure defines how to assign variables into various groups in the penalty term $\Omega$. The objective of pairing a regularization method with a grouping structure is to implement restrictions on the combinations of variables that can be selected together (which corresponds to respecting a selection rule). The actual variables selected in an analysis will depend on the data, $D$.  Similar to the selection rule dictionary, we define the penalization structure dictionary below.
\begin{definition}[Penalization structure dictionary]
\label{def.PSD}
Given a method $\mathcal{M}$, and a compatible grouping structure $\mathbb{G}$ on $\mathbb{V}$, there is one corresponding \textbf{penalization structure dictionary} $\mathbb{D}^{\mathcal{M}}_{\mathbb{G}}$, which is a 
dictionary that contains all subsets of $\mathbb{V}$ that could potentially result from the application of the penalization structure on sample data.
\end{definition}
\noindent
Similar to the relationship between selection rules and selection rule dictionaries, we say the resulting penalization structure dictionary is congruent to the penalization structure, denoted by $\{\mathcal{M},\mathbb{G}\}\cong\mathbb{D}_{\mathbb{G}}^{\mathcal{M}}$. The next theorem reveals the connection between the selection rule and penalized regression through their dictionaries. It can be conceptually presented in Figure \ref{fig:Concept}.
\begin{theorem}\label{th:GeneralTheorem}
    When the penalization structure dictionary $\mathbb{D}_{\mathbb{G}}^{\mathcal{M}}$ equals a selection dictionary $\mathbb{D}_{\mathfrak{r}}$ of a rule $\mathfrak{r}$, we say that the penalization structure $\{\mathcal{M},\mathbb{G}\}$ respects the selection rule $\mathfrak{r}$. That is, the sufficient and necessary condition of the penalization structure $\{\mathcal{M},\mathbb{G}\}$ being congruent to the selection rule $\mathfrak{r}$ is $\{\mathcal{M},\mathbb{G}\}\cong\mathbb{D}_{\mathbb{G}}^{\mathcal{M}}=\mathbb{D}_{\mathfrak{r}}\cong\mathfrak{r}$, . 
\end{theorem}
The above theorem is a direct result of the uniqueness of the selection dictionary and the penalization structure dictionary (with regard to the congruent selection rule and penalization structure). In practice, given a selection rule and the derived selection dictionary, we aim to find the grouping structure with a penalization regression whose penalization structure dictionary equals the selection dictionary so that the selection rule can be respected.

In addition, the grouping structure can be shown graphically, where the variables in a same group $\mathbb{g}$ are in the same closed curve. We use some examples to illustrate various grouping structures with different regularization methods. 

Suppose $\mathbb{V}=\{A, B, C, D\}$, $\mathbb{F}_1=\{A, B\}$ and $\mathbb{F}_2=\{C, D\}$, and let the  grouping structures of Examples 1-3 be $\mathbb{G}=\{\mathbb{F}_1, \mathbb{F}_2\}$.\\
\textbf{Example 1} Group Lasso \citep{yuan2006model} can achieve the groupwise selection  $\mathfrak{r}=\mathfrak{u}_{\{0, 2\}}(\mathbb{F}_{1})\land\mathfrak{u}_{\{0, 2\}}(\mathbb{F}_{2})$.\\
\textbf{Example 2} Exclusive group Lasso \citep{campbell2017within} can achieve the uncorrelated (or within) group selection $\mathfrak{r}=\mathfrak{u}_{\{1, 2\}}(\mathbb{F}_{1})\land \mathfrak{u}_{\{1, 2\}}(\mathbb{F}_{2})$.  \\
\textbf{Example 3} Sparse group Lasso \citep{simon2013sparse} can achieve the bi-level selection $\mathfrak{u}_{\{0, 1, 2\}}(\mathbb{F}_{1})\land \mathfrak{u}_{\{0, 1, 2\}}(\mathbb{F}_{2})$. \\
We see that a same grouping structure with different regularization methods can respect different selection rules. However, one penalization structure can only respect one selection rule.

The grouping structures outlined above are straightforward because of the inherent design of penalized regression and the simple selection rules. Nevertheless, it is important to note that each method is limited in its ability to accommodate a specific type of selection. In contrast, certain other techniques, such as the LOGL, exhibit the capacity to accommodate more complicated selection rules. However, the challenge lies in effectively arranging variables into groups to achieve the desired dictionary. In the following sections, we shift our focus towards LOGL and offer guidance on the identification of grouping structures, with the aim of enabling the incorporation of a broader range of selection rules.

\section{More details of the roadmaps}
\label{App:roadmap}
More than one grouping structure can respect a given selection rule under latent overlapping group lasso. The roadmap provided in Table \ref{tab:roadmap} shows the way to identify the most efficient grouping structure in the sense that the grouping structure contains the least number of groups among all eligible grouping structures.\\
Within the interaction selection application, when all the variables in $\mathbb{A}$ are the interactions of the variables in $\mathbb{B}$, selection rules 1 and 2 in Table \ref{tab:roadmap} correspond to strong and weak heredity, respectively. \citep{Haris_2016} When $n=1$ and $m=2$, the two selection rules degrade to two-way interaction selection. In this simple case, the grouping structure construction was given by \citep{yan2017hierarchical}.\\
Note that the roadmap works with conditions: all variables in $\mathbb{A}$ and $\mathbb{B}$ are continuous or binary variables, and $\mathbb{A}\cap\mathbb{B}=\emptyset$.
When there are categorical variables in $\mathbb{B}$, and  thus $\mathbb{A}$, one should be more careful to specify the grouping structure in the sense that it is necessary to group the dummy variables in $\mathbb{B}$ representing a categorical variable, but not the ones in $\mathbb{A}$. In addition, grouping structures must be specified case by case when there is more than one rule being applied to a set of variables. For example, suppose that we want to respect the two selection rules ``if A is selected, then \{B, C\} must be selected'' and ``if D is selected, then $\{A, B, C\}$ must be selected''. Then the grouping structure should be $\big\{\{B\}, \{C\}, \{A, B, C\}, \{A, B, C, D\}\big\}$. $\{A\}$ should not be a single group because it cannot be selected alone.\\ 

\section{Rationale of the modification of rule 3}
\label{App:rule3}
The latent overlapping group lasso cannot respect the selection rule 3, which states that all of the 10 variables must be selected. This is because, first, the latent overlapping group lasso requires that all variables must belong to at least one group in the penalty term: all coefficients of variables must be penalized at a certain level. Once a group is specified, the group of variables are eligible to not be selected. Since the weight $\omega_{\mathbb{g}}$ must be positive, one may attempt to set the weight of the group to a fairly small number close to 0, which is effectively the same as no penalization on the variables contained in the group. However, the estimates of the latent overlapping group lasso are very sensitive to the weight specification. A zero weight for a group is not allowed, see details in \citep{obozinski2011variable}.\\ 
We set the rule 3 because the literature has established the relevance of these 10 variables and so no model would be accepted by subject matter experts without them. Given that we expect these variables to be highly predictive of our outcome, the modified rule (select the 10 variables as a group) is pragmatically the same as the original selection rule, because it's effectively impossible that none of these variables would be important for prediction.

\section{Grouping structure in the application}
\label{GS_app}
Define\\ 
$\mathbb{g}_{1}$=\{CHA$_{2}$DS$_{2}$-VASc Score\},\\
$\mathbb{g}_{2}$=\{Heart Disease\},\\
$\mathbb{g}_{3}$=\{Diabetes\},\\
$\mathbb{g}_{4}$=\{COPD\},\\
$\mathbb{g}_{5}$=\{Dyslipidemia\},\\
$\mathbb{g}_{6}$=\{DOAC\},\\
$\mathbb{g}_{7}$=\{DOAC, High-dose-DOAC\},\\
$\mathbb{g}_{8}$=\{DOAC, Apixaban\},\\
$\mathbb{g}_{9}$=\{DOAC, Dabigatran\},\\
$\mathbb{g}_{10}$=\{High-adherence\},\\
$\mathbb{g}_{11}$=\{DOAC, High-adherence, Interaction of DOAC and High-adherence\},\\
$\mathbb{g}_{12}$=\{DOAC, High-adherence, Interaction of DOAC and High-adherence, High-dose-DOAC, Interaction of High-dose-DOAC and High-adherence\},\\
$\mathbb{g}_{13}$=\{DOAC, Interaction of DOAC and Antiplatelets\},\\
$\mathbb{g}_{14}$=\{DOAC, Interaction of DOAC and NSAIDs\},\\
$\mathbb{g}_{15}$=\{Antidepressants\},\\
$\mathbb{g}_{16}$=\{DOAC, Antidepressants, Interaction of DOAC and Antidepressants\},
\\$\mathbb{g}_{17}$=\{PPI\},\\
$\mathbb{g}_{18}$=\{DOAC, PPI, Interaction of DOAC and PPI\},\\
$\mathbb{g}_{19}$=\{Age, History of major bleeding, Stroke, Anemia, Sex, Renal diseases, Liver disease, Malignancy, Antiplatelets, NSAIDs\}.\\
The grouping structure is $\mathbb{G}=\{\mathbb{g}_{i}, i=1,\dots,19\}$. 

Note that there is no single-variable group for the variable Antiplatelets or NSAIDs, as it should be for respecting rules 2.3 and 2.4. This is because these two variables are forced to be selected by rule 3.

\section{Descriptive statistics of each covariate}\label{appendix:TableOneVariables}
\begin{longtable}{p{10cm}ll}
\hline
\textbf{Variable name (non-reference/reference level)} &  \textbf{Non-bleeding} & \textbf{Bleeding}\\
& n=20,671 (96.90\%) &n=661 (3.10\%)\\
\hline\hline
\multicolumn{3}{c}{\textbf{Baseline covariates}}\\
1. Age, Mean (SD) & 79.98 (8.79) & 81.49 (7.93)\\
2. Sex (proportion female) & 0.58 & 0.54\\
3. CHA$_{2}$DS$_{2}$-VASc Score ($\geqslant$3/$<$3) & 0.88  & 0.92\\
\multicolumn{3}{c}{\textbf{Comorbidities within 3 years before cohort entry}}\\
4. Stroke(yes/no)& 0.28& 0.26 \\
5. Anemia (yes/no) & 0.10  & 0.17 \\
6. Malignancy (yes/no) & 0.26  & 0.30 \\
7. Liver disease (yes/no) & 0.02  & 0.05\\
8. History of major bleeding (yes/no)& 0.34  & 0.49 \\
9. Renal diseases (yes/no)&0.27  &0.34  \\
10. Heart disease(yes/no) & 0.63  & 0.72 \\
11. Diabetes (yes/no) & 0.36  & 0.46 \\
12. COPD/asthm (yes/no)& 0.39  & 0.46 \\
13. Dyslipidemia (yes/no)& 0.58  & 0.64 \\
\multicolumn{3}{c}{\textbf{OAC use at cohort entry}}\\
14. DOAC (DOACs/warfarin)  & 0.57  &0.50\\
15. Apixaban (yes/no) &  0.30 & 0.21 \\
16. Dabigatran (yes/no) & 0.10 & 0.12 \\
17. High-dose-DOAC (high-dose-DOACs/low-dose-DOACs or warfarin) & 0.33 & 0.25\\
18. High-adherence (high/low)   & 0.87 & 0.25 \\
19. Interaction of DOAC and High-adherence  & 0.49  & 0.43 \\
20. Interaction of High-dose-DOAC and High-adherence  & 0.28  & 0.22 \\
\multicolumn{3}{c}{\textbf{Concomitant medication use within 2 weeks before cohort entry}}\\
21. Antiplatelets (yes/no) & 0.32  & 0.40 \\
22. NSAIDs (yes/no)& 0.01 & 0.01 \\
23. Antidepressants (yes/no)& 0.18 & 0.22 \\
24. PPIs (yes/no)& 0.44 & 0.44 \\
\multicolumn{3}{c}{\textbf{Potential drug-drug interaction*}}\\
25. Interaction of DOAC and Antiplatelets & 0.16  & 0.16 \\
26. Interaction of DOAC and NSAIDs & 0.01  & 0.01\\
27. Interaction of DOAC and Antidepressants & 0.10 & 0.10\\
28. Interaction of DOAC and PPIs & 0.23 & 0.19 \\
\hline
\caption{Covariate descriptive statistics (prevalences for binary covariates and means and standard errors for continuous covariates) stratified by the outcome}
\end{longtable}
\section{Demographic and characteristics of patients stratified by DOAC dose and warfarin}
\label{SSSD}
\footnotesize
\begin{longtable}{p{7.5cm}lll}
\hline
& \multicolumn{2}{c}{\textbf{DOAC}}\\\cmidrule(lr){2-4}
\textbf{Variable name} &  \textbf{High-dose} & \textbf{Low-dose} & \textbf{warfarin}\\
\textbf{(non-reference/reference level)}\\
& n=7022 (32\%) &n=5067 (24\%)&n=9243 (43\%)\\
\hline\hline
\multicolumn{4}{c}{\textbf{Baseline covariates}}\\
1. Age ($\geqslant$75/$<$75) & 0.55 &  0.91 & 0.77 \\
2. Sex (female/male) & 0.50  & 0.66  & 0.59 \\
3. CHA$_{2}$DS$_{2}$-VASc Score ($\geqslant$3/$<$3) & 0.79 & 0.95&   0.91\\
\multicolumn{4}{c}{\textbf{Comorbidities within 3 years before cohort entry}}\\
4. Stroke (yes/no)& 0.25  &  0.27  &  0.30 \\
5. Anemia (yes/no) & 0.06& 0.09  &  0.13 \\
6. Malignancy (yes/no) & 0.27 &  0.26   & 0.26 \\
7. Liver disease (yes/no) &  0.02  &  0.02  & 0.02 \\
8. History of major bleeding (yes/no)& 0.28  &  0.36 &  0.37 \\
9. Renal diseases (yes/no) &0.18&  0.27  & 0.35  \\
10. Heart disease (yes/no) & 0.55  & 0.65   & 0.69 \\
11. Diabetes (yes/no) & 0.36 &  0.30 &  0.40 \\
12. COPD/asthma (yes/no)& 0.38  &  0.37  & 0.41 \\
13. Dyslipidemia (yes/no)& 0.59  & 0.55   & 0.59 \\
\multicolumn{4}{c}{\textbf{OAC use at cohort entry}}\\
14. DOAC (DOAC/warfarin)  & 1.00   & 1.00  &  0.00\\
15. Apixaban &0.54 &  0.50  &  0.00\\
16. Dabigatran &  0.11  & 0.28& 0.00 \\
17. High-dose-DOAC (high-dose-DOAC/low-dose-DOAC or warfarin) & 1.00  &  0.00  &  0.00 \\
18. High-adherence (high/low) & 0.85  & 0.86  & 0.88 \\
19. Interaction of DOAC and High-adherence & 0.85  & 0.86  & 0.00 \\
20. Interaction of High-dose-DOAC and High-adherence  & 0.85  & 0.00  & 0.00 \\
\multicolumn{4}{c}{\textbf{Concomitant medication use within 2 weeks before cohort entry}}\\
21. Antiplatelets (yes/no) & 0.27  &0.32  &  0.37 \\
22. NSAIDs (yes/no)& 0.01  & 0.01  &  0.01 \\
23. Antidepressants (yes/no)& 0.18   & 0.20   & 0.18 \\
24. PPIs (Proton pump inhibitors) (yes/no)& 0.37   & 0.44  & 0.49\\
\multicolumn{4}{c}{\textbf{Potential drug-drug interaction}}\\
25. Interaction of DOAC and Antiplatelets & 0.27  & 0.32  & 0.00 \\
26. Interaction of DOAC and NSAIDs & 0.01 & 0.01 &   0.00\\
27. Interaction of DOAC and Antidepressants & 0.18 & 0.20 & 0.00 \\
28. Interaction of DOAC and PPIs & 0.37  & 0.44 &  0.00 \\
\multicolumn{4}{c}{\textbf{Outcome}}\\
29. Major bleeding & 0.02   & 0.03 & 0.04 \\
\hline
\caption{Covariates proportions stratified by Dose}
\label{TableOneVariables_DOACdose}%
\end{longtable}
\normalsize

\section{Crude and adjusted odds ratios and 95\% confidence intervals of variables}
\label{Crud_Adjusted}
\begin{table}[ht]
    \begin{center}
    \begin{tabular}{p{7cm}ll}\hline
        \textbf{ Variable name (non-reference/reference level)}      &  \textbf{ Crude OR (95\% CI)} & \textbf{ Adjusted OR (95\% CI)}\\\hline\hline
    Elderly ($>$65/$\leqslant$ 65) & 2.51  (1.53, 4.48) & 2.43 (1.48, 4.35)\\
    History of major bleeding (yes/no) & 1.87  (1.60, 2.19)   & 1.78 (1.51, 2.08)\\
    Liver diseases (yes/no)& 2.35 (1.60, 3.34) & 2.15 (1.46, 3.06)\\
    Renal diseases (yes/no) & 1.36 (1.15, 1.60) & 1.14  (0.96, 1.34)\\
    Stroke (yes/no)& 0.96  (0.85, 1.07) & 0.93  (0.82, 1.04)\\
    Drugs* (yes/no)& 1.23  (1.04, 1.45) & 1.18  (1.00, 1.40)\\ \hline
    \end{tabular}
    \caption{Crude (univariate) and adjusted odds ratios from simple and multivariate logistic regression models respectively and 95\% confidence intervals using the variables in HAS-BLED chart (2012, Circulation, Lane at el.)}
    \end{center}
    {\footnotesize* Drugs: a binary variable, it is 1 if the patient had at least 1 dispensation of the following drugs in the 3 years preceding the index date or during the atrial fibrillation hospitalization: Clopidogrel, low-dose of ASA (daily dose $<$ 100 mg), NSAID. }
    \label{tab:has-bled}
\end{table}

\begin{table}[ht]
    \begin{center}
    \begin{tabular}{p{7cm}ll}\hline
        \textbf{ Variable name (non-reference/reference level)}      &  \textbf{ Crude OR (95\% CI)} & \textbf{ Adjusted OR (95\% CI)}\\\hline\hline
Age ($\geqslant$75/$<$75)&   1.28  (1.07,   1.55)&  1.29  (1.07,   1.56)\\
Sex (female/male)&   0.84  (0.72,   0.98)&  0.86  (0.73,   1.01)\\
Adherence (high/low)&   1.04  (0.83,   1.32)& 1.03  (0.82,   1.32) \\
History of major bleeding (yes/no)&   1.87  (1.60,   2.19)&  1.68  (1.43,   1.98)\\
Renal diseases (yes/no)&   1.36  (1.15,   1.60)&  1.05  (0.88,   1.25)\\
Liver diseases (yes/no)&  2.35  (1.60,   3.34) &  2.09  (1.42,   2.99)\\
Stroke (yes/no)&   0.96  (0.85,   1.07)&  0.94  (0.83,   1.05)\\
Anemia (yes/no)&   1.86  (1.50,   2.28)&  1.46  (1.16,   1.81)\\
Malignancy (yes/no) &  1.21  (1.02,   1.43) &  1.09  (0.91,   1.29)\\
Antiplatelets (yes/no)&  1.40  (1.19,   1.64) &  1.31  (1.11,   1.54)\\
NSAIDs (yes/no)&  1.15  (0.54,   2.11) &  1.29  (0.61,   2.38)\\
Other medications* (yes/no) &  1.22  (1.03,   1.46) &  1.06  (0.88,   1.27)
\\ \hline
    \end{tabular}
    \caption{Crude (univariate) and adjusted odds ratios from simple and multivariate logistic regression models respectively and 95\% confidence intervals using the risk factors of bleeding on Oral anticoagulation  (2012, Circulation, Lane at el.)}
    \end{center}
    {\footnotesize Other medications: a binary variable, it is 1 if the patient had at least 1 dispensation of the following drugs within the 2 weeks prior to the index date: Antidiabetics, Antidepressants, PPIs, Antibiotics, Antiarrhythmics, PGP inhibitors.}
    \label{tab:risk-factor}
\end{table}
\clearpage

\footnotesize
\begin{longtable}{p{8.5cm}ll}\hline
    \textbf{ Variable name (non-reference/reference level)}      &  \textbf{ Crude OR (95\% CI)} & \textbf{ Adjusted OR (95\% CI)}\\\hline\hline
    \multicolumn{3}{c}{\textbf{Baseline covariates}}\\
1. Age ($\geqslant$75/$<$75)&   1.28  (1.07,   1.55)  &   1.23  (1.00,   1.54)\\
2. Sex (female/male)&    0.84  (0.72,   0.98) &  0.86  (0.73,   1.02)\\
3. CHA$_{2}$DS$_{2}$VASc score ($\geqslant$3/$<$3)&  1.54  (1.17,   2.06)   & 1.24  (0.90,   1.75) \\
\multicolumn{3}{c}{\textbf{Comorbidities within 3 years before cohort entry}}\\
4. Stroke (yes/no)&   0.96  (0.85,   1.07)  &   0.94  (0.83,   1.05)\\
5. Anemia (yes/no)&  1.86  (1.50,   2.28)  &   1.38  (1.10,   1.72)\\
6. Malignancy (yes/no) &    1.21  (1.02,   1.43) &  1.09  (0.91,   1.30)\\
7. Liver diseases (yes/no) (yes/no)&   2.35  (1.60,   3.34) &   2.01  (1.36,   2.89)\\
8. History of major bleeding (yes/no)& 1.87  (1.60,   2.19)   &  1.62  (1.37,   1.91) \\
9. Renal diseases (yes/no)&   1.36  (1.15,   1.60)  & 0.96  (0.80,   1.15) \\
10. Heart disease (yes/no)&   1.51  (1.27,   1.80)  &  1.23  (1.02,   1.48)\\
11. Diabetes (yes/no)&  1.49  (1.27,   1.74)  &  1.30  (1.10,   1.54)\\
12. COPD/asthma (yes/no)&   1.33  (1.14,   1.55)  &   1.15  (0.98,   1.35)\\
13. Dyslipidemia (yes/no)&  1.28  (1.09,   1.50)  &   1.09  (0.92,   1.30)\\
\multicolumn{3}{c}{\textbf{OAC use at cohort entry}}\\
14. DOAC (DOACs/warfarin)&  0.76  (0.65,   0.89)  &  1.69  (0.94,   2.98)\\
15. Apixaban (yes/no), ref: Rivaroxaban &  0.63  (0.52,   0.76)  & 0.64  (0.50,   0.82) \\
16. Dabigatran (yes/no), ref: Rivaroxaban &  1.17  (0.91,   1.47)  &  1.07  (0.78,   1.45)\\
17. High-dose-DOAC (high-dose-DOACs/low-dose-DOACs or warfarin)&  0.68  (0.56,   0.81)  &  0.66  (0.36,   1.19)\\
18. High-adherence (high/low)&   1.04  (0.83,   1.32) & 1.02  (0.72,   1.48) \\
19. Interaction of DOAC and High-adherence&   0.79  (0.67,   0.92) &  0.82  (0.47,   1.45)\\
20. Interaction of High-dose-DOAC and High-adherence&  0.71  (0.58,   0.85)  &  1.29  (0.69,   2.42)\\
\multicolumn{3}{c}{\textbf{Concomitant medication use within 2 weeks before cohort entry}}\\
21. Antiplatelets (yes/no)&   1.40  (1.19,   1.64) &   1.48  (1.18,   1.86)\\
22. NSAIDs (yes/no)&  1.15  (0.54,   2.11)  & 1.03  (0.25,   2.78) \\
23. Antidepressants (yes/no)&  1.27  (1.05,   1.52)  & 1.22  (0.93,   1.59) \\
24. PPIs (yes/no)&   1.03  (0.88,   1.20) &  0.85  (0.68,   1.07)\\
\multicolumn{3}{c}{\textbf{Potential drug-drug interaction}}\\
25. Interaction of DOAC and Antiplatelets&  0.95  (0.76,   1.17)  &  0.71  (0.51,   0.99)\\
26. Interaction of DOAC and NSAIDs&   1.26  (0.49,   2.62)  &  1.52  (0.38,   7.42)\\
27. Interaction of DOAC and Antidepressants&   0.98  (0.75,   1.26)  &  0.90  (0.61,   1.32)\\
28. Interaction of DOAC and PPIs&  0.78  (0.64,   0.95)  &  0.88  (0.63,   1.22)
\\ \hline
    \caption{Crude (univariate) and adjusted odds ratios from simple and multivariate logistic regression models, respectively and 95\% confidence intervals using the covariates in in the analysis}
    \label{tab:crud-adj_analysis}
\end{longtable}
\normalsize
\section{Derivation of estimated odds ratios of taking different types of DOACs versus warfarin from the selected models}
\label{rationale}
We focus on the selected model resulting from the latent overlapping group lasso using MCP/SCAD penalty. Based on the current variable definitions, we show the corresponding variable values when the patient took different types of OACs below.\\
\begin{table}[h]
    \centering
    \begin{tabular}{ccccc}
    \hline
    Category/Variable name     &  DOAC & High-dose-DOAC & Apixaban & Dabigatran\\\hline\hline
    High-dose-Apixaban      & 1 & 1& 1& 0\\
    High-dose-Dabigatran     & 1 & 1& 0& 1\\
    High-dose-Rivaroxaban & 1 & 1& 0& 0\\
    Low-dose-Apixaban   & 1 & 0& 1& 0\\
    Low-dose-Dabigatran     & 1 & 0& 0& 1\\
    Low-dose-Rivaroxaban     & 1 & 0& 0& 0\\
    warfarin     & 0 & 0& 0& 0\\\hline
    \end{tabular}
    \caption{Variable values when the patient took different types of OACs (category)}
    \label{tab:my_label}
\end{table}\\
\noindent
Under the logistic regression, we have 
$$
    logit\{Pr(Y=1\mid \mathbf{X}, \boldsymbol{\beta})\}=\beta_0 +\beta_1 \text{DOAC }+\beta_2 \text{High-dose-DOAC} +\beta_3\text{Apixaban}+\beta_4 \text{Dabigatran}+...,
$$
where $Y$ is the outcome, major bleeding. For brevity, we omit the other variables. Then the combinations of coefficients used to estimate the probabilities of major bleeding for different types of OACs are given in Table \ref{tab:my_label} and correspondingly with notation in Table \ref{dss}. The estimates of the odds ratios contrasting each type of DOAC usage with warfarin are given in Table~\ref{ddd}\\
\begin{table}[h]
    \centering
    \begin{tabular}{cc}
    \hline
    Subject who took...    & Estimates \\
    \hline\hline
   High-dose-Apixaban      & ilogit($\hat{\beta}_0+\hat{\beta}_1+\hat{\beta}_2+\hat{\beta}_3+\dots$)\\
    High-dose-Dabigatran      & ilogit($\hat{\beta}_0+\hat{\beta}_1+\hat{\beta}_2+\hat{\beta}_4+\dots$)\\
   High-dose-Rivaroxaban       & ilogit($\hat{\beta}_0+\hat{\beta}_1+\hat{\beta}_2+\dots$)\\
    Low-dose-Apixaban    & ilogit($\hat{\beta}_0+\hat{\beta}_1+\hat{\beta}_3+\dots$)\\
    Low-dose-Dabigatran   & ilogit($\hat{\beta}_0+\hat{\beta}_1+\hat{\beta}_4+\dots$)\\
    Low-dose-Rivaroxaban    & ilogit($\hat{\beta}_0+\hat{\beta}_1+\dots$)\\
    warfarin     & ilogit($\hat{\beta}_0+\dots$)\\\hline
    \end{tabular}
    \caption{Estimated mean outcomes}
    \label{dss}
    *ilogit: inverse logit
\end{table}\\
\noindent

\begin{table}[h]
    \centering
    \begin{tabular}{ccc}
    \hline
    Contrasts   &  Parameter(s) & Estimated odds ratios\\
    \hline\hline
    High-dose-Apixaban    & exp($\hat{\beta}_1+\hat{\beta}_2+\hat{\beta}_3$) & 0.70\\
    High-dose-Dabigatran     & exp($\hat{\beta}_1+\hat{\beta}_2+\hat{\beta}_4$) & 1.13\\
    High-dose-Rivaroxaban     & exp($\hat{\beta}_1+\hat{\beta}_2$) & 1.13 \\
    Low-dose-Apixaban    & exp($\hat{\beta}_1+\hat{\beta}_3$) & 0.86\\
    Low-dose-Dabigatran    & exp($\hat{\beta}_1+\hat{\beta}_4$) & 1.39\\
    Low-dose-Rivaroxaban   & exp($\hat{\beta}_1$) & 1.39\\
   \hline
\hline
    \end{tabular}
    \caption{Estimated odds ratios of taking different types of DOACs versus warfarin from the model selected by the latent overlapping group lasso MCP/SCAD}
    \label{ddd}
\end{table}
\end{document}